\documentclass[11pt]{article}
\pdfoutput=1

\usepackage{jheppub} 
\usepackage{graphicx}
\usepackage{dcolumn}
\usepackage{bm}

\usepackage{mathtools}
\usepackage{amsmath}
\usepackage{amssymb}
\usepackage{braket}
\usepackage{color}
\usepackage{xcolor}
\usepackage{hyperref}
\usepackage{slashed}
\usepackage{graphicx}
\usepackage{mathrsfs}
\usepackage{tikz}
\usepackage{tikz-cd}
\usepackage{enumitem}
\usepackage{eso-pic}

\usepackage{mathtools} 
\usepackage{extarrows} 
\usepackage{multirow}

\usepackage{tikz}
\usepackage{diagbox}
\usetikzlibrary{positioning}
\usetikzlibrary{calc}
\usetikzlibrary{decorations.pathreplacing,calligraphy}

\usepackage{xstring}
\usetikzlibrary{decorations.pathmorphing} 
\usetikzlibrary{decorations.markings} 
\usetikzlibrary{arrows} 
\usetikzlibrary{shapes} 
\usetikzlibrary{matrix} 
\usetikzlibrary{positioning} 
\usepackage[english]{babel} 
\usepackage[autostyle]{csquotes}
\usepackage{subfig}

\tikzstyle{brane}=[draw]
\tikzset{D7/.style={circle, draw=black, inner sep=0pt, fill=white, minimum size=3mm}}
\tikzset{hasse/.style={circle, fill,inner sep=2pt}}
\tikzset{flavor/.style={regular polygon,fill=white,regular polygon sides=4,inner sep=2.5pt, draw}}
\tikzset{gauge/.style={circle, draw,inner sep=2.5pt}}
\tikzset{gaugeb/.style={circle, draw,fill=black,inner sep=2.5pt}}
\tikzset{gauger/.style={circle, draw,fill=cyan,inner sep=2.5pt}}
\tikzset{gaugeg/.style={circle, draw,fill=red,inner sep=2.5pt}}
\tikzset{bd/.style={circle, draw=black, inner sep=0pt, fill=black, minimum size=2mm}}
\tikzset{wd/.style={circle, draw=black, inner sep=0pt, fill=white, minimum size=2mm}}
\tikzset{Dynkin/.style={circle, draw=black, inner sep=0pt, fill=white, minimum size=2mm}}
\tikzstyle{ligne}=[draw, thick] 
\tikzset{doublearrow/.style={ draw=black!75, color=black!75, thick, double distance=3pt, }}

\numberwithin{equation}{section}  

\graphicspath{ {figs/} }

\newcommand{\be}{\begin{equation}}
\newcommand{\ee}{\end{equation}}
\newcommand{\ba}{\begin{aligned}}
\newcommand{\ea}{\end{aligned}}

\def\unit{{1\kern-.65ex {\rm l}}}
\def\1{{1\kern-.65ex {\rm l}}}

\newcount\hour \newcount\minute
\hour=\time \divide \hour by 60
\minute=\time
\def\now{%
\ifnum \hour<13
  \ifnum \hour=0 \advance \hour by 12 \number\hour:\else \number\hour:\fi%
     \ifnum \minute<10 0\fi%
     \number\minute%
\ A.M.%
\else \advance \hour by -12 \number\hour:%
  \ifnum \minute<10 0\fi%
  \number\minute%
  \ P.M.%
\fi%
}

\makeatother

\def\mb{\mathbb}
\def\mbf{\mathbf}
\def\mc{\mathcal}

\def\bp{\begin{pmatrix}}
\def\ep{\end{pmatrix}}

\def\ptl{\partial}

\newcommand{\bea}{\begin{equation} \begin{aligned}}
 \newcommand{\eea}{\end{aligned} \end{equation}}
\newcommand{\bit}{\begin{itemize}} 
\newcommand{\eit}{\end{itemize}}

\renewcommand{\d}{\partial }

\tikzstyle{brane}=[draw]
\tikzset{D7/.style={circle, draw=black, inner sep=0pt, fill=white, minimum size=3mm}}
\tikzset{hasse/.style={circle, fill,inner sep=2pt}}
\tikzset{flavor/.style={regular polygon,fill=white,regular polygon sides=4,inner sep=2.5pt, draw}}
\tikzset{gauge/.style={circle, draw,inner sep=2.5pt}}
\tikzset{gaugeb/.style={circle, draw,fill=black,inner sep=2.5pt}}
\tikzset{gauger/.style={circle, draw,fill=cyan,inner sep=2.5pt}}
\tikzset{gaugeg/.style={circle, draw,fill=red,inner sep=2.5pt}}
\tikzset{SUd/.style={circle, draw=black, inner sep=0pt, fill=yellow, minimum size=2mm}}
\tikzset{bd/.style={circle, draw=black, inner sep=0pt, fill=black, minimum size=2mm}}
\tikzset{wd/.style={circle, draw=black, inner sep=0pt, fill=white, minimum size=2mm}}
\tikzset{Dynkin/.style={circle, draw=black, inner sep=0pt, fill=white, minimum size=2mm}}
\tikzstyle{ligne}=[draw, thick] 
\tikzset{doublearrow/.style={ draw=black!75, color=black!75, thick, double distance=3pt, }}

\usepackage{booktabs}
\usepackage{makecell}

\def\ms{\mathscr}
\newcommand{\dd}{\mathrm{d}}
\renewcommand{\d}{\partial }

\newcommand{\op}[1]{\operatorname{#1}}

\tikzset{global scale/.style={
    scale=#1,
    every node/.append style={scale=#1}
  }
}

\begin{document}

\thispagestyle{empty}

\vspace*{0.8cm} 
\begin{center}
{{\Huge  Vacuum Tunneling from Conifold Transitions in IIB
}}

\vspace*{1.5cm}

 {\large Xin Gao$^1$, Qinjian Lou$^2$, Yi-Nan Wang$^{2,3}$}

\vspace*{1.0cm}

\smallskip

{\it $^1$College of Physics, Sichuan University,  \protect\\
Chengdu, 610065, China}

{\it $^2$School of Physics, Peking University, \protect\\
Beijing 100871, China}

{\it $^3$Center for High Energy Physics, Peking University, \protect\\
Beijing 100871, China}

\smallskip

\vspace*{0.8cm}
\end{center}

\vspace*{.5cm}

\noindent

We investigate the quantum tunneling process through a phase transition near a conifold singularity, in the setup of IIB CY3 orientifold compactification. We propose a novel method to do moduli stabilization in an extended moduli space, parametrized by both the geometric moduli and the light D3-brane wrapping modes arising from the brane quantization. Assuming the absence of flux through the vanishing exceptional 3-cycle, we find two types of vacuum solutions, one corresponds to the resolved conifold and the other one is interpreted as a novel non-geometric phase. We provide an estimation on the quantum tunneling rate between these two solutions and find that it is difficult to achieve a significantly large tunneling rate in the controllable regime.

\newpage

\tableofcontents

\newpage

\flushbottom

\section{Introduction}

It is known that the different ways of compactifying superstring/M/F-theory to 4-dimensional spacetime giving rise to a vast string landscape of metastable vacua, see e.g.~\cite{Lerche:1986cx,Denef:2004ze,Douglas:2006es,Denef:2008wq,Taylor:2011wt,Taylor:2015xtz,Brennan:2017rbf,Cvetic:2019gnh,Hebecker:2020aqr,Demirtas:2020dbm}. The construction of a geometric landscape of vacua shall be achieved by the following steps, after choosing a particular version of string theory:
\begin{enumerate}
\item Choosing the topology of the compact manifold $X$.

\item Choosing a configuration of brane objects in the geometric background, e.g. 7-branes in F-theory or orientifold branes and D-branes in weakly coupled string theory setups. 

\item Choosing discrete $p$-form flux, e.g. $G_4$ flux in M/F-theory or $F_3,H_3$ flux in IIB.

\item Compute the scalar potential $V$ under the aforementioned choices, and stabilize the moduli fields, to get a set of metastable vacua.
\end{enumerate}

Assuming we live in (or near) a metastable vacuum, it is natural to wonder what are the physical roles of other vacua associated with different flux choices, D-brane configurations and even different topologies of compact manifolds\footnote{We do not attempt to solve the problem of constructing de Sitter vacua in string theory (see the recent progress and reviews in e.g. \cite{Cicoli:2018kdo,Dasgupta:2019gcd,Crino:2020qwk,Berglund:2022qsb,McAllister:2024lnt,Burgess:2024jkx}), and we are only interested in the difference of potential $\delta V$ for two local minima of the scalar potential $V$.}.

\begin{figure}[h]
\centering
\includegraphics[height=6cm]{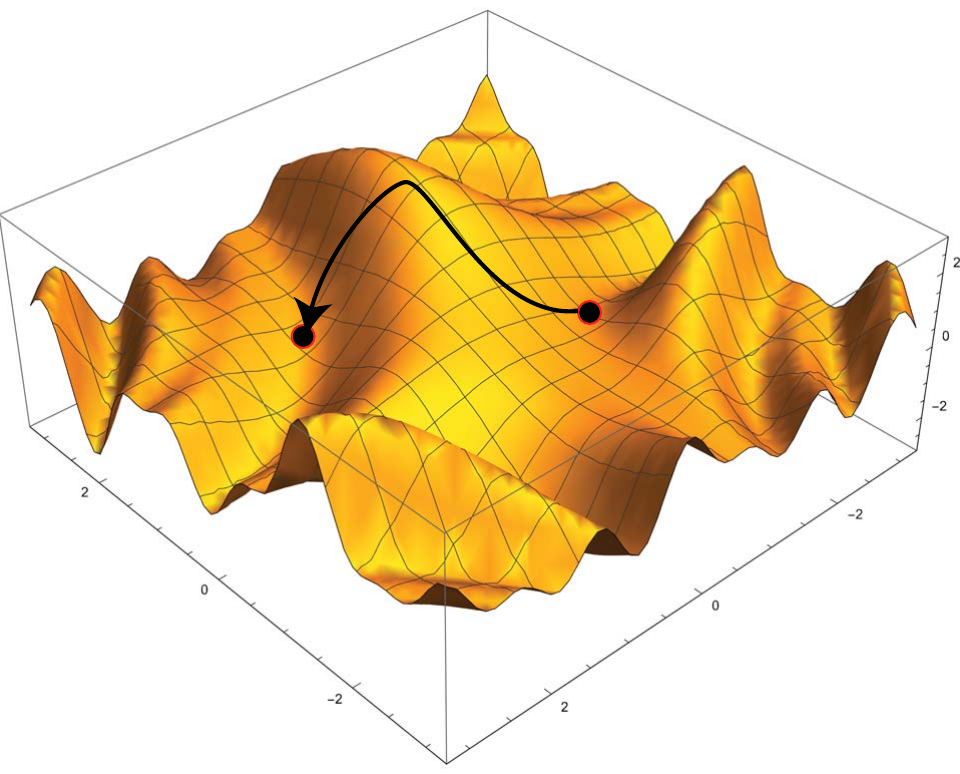}
\caption{A demonstration of quantum tunneling between two local minima in the string landscape.}\label{f:landscape}
\end{figure}

In this paper, we investigate the possible physical implications of the existence of multiple vacua in the string landscape, from the quantum tunneling process between two local minima of $V$, i.e. figure~\ref{f:landscape}. In particular, we formulate a new framework to investigate the quantum tunneling between manifolds with different topologies, extending the previously studied quantum tunneling effect between different flux choices~\cite{Johnson:2008kc,Blanco-Pillado:2009lan,Ahlqvist:2010ki,deAlwis:2013gka}.

The idea is to introduce an extended moduli space of the compact space: $\mc{M}_E$, whose lower-dimensional subsets $\mc{M}_1\subset\mc{M}_E$, $\mc{M}_2\subset\mc{M}_E$ describe compact spaces with different topologies. The intersection locus $\mc{M}_1\cap\mc{M}_2$ precisely describes the topology transition point. For the cases of Calabi-Yau threefolds (CY3), $\mc{M}_E$ can be as large as the space of all compact CY3-folds, under the principles of Reid's fantasy~\cite{reid1987moduli}. Nonetheless, in this paper we will only investigate a tiny fraction of $\mc{M}_E$, consists of two compact CY3-folds $\mc{X}_6$ with $(h^{1,1}(\mc{X}_6),h^{2,1}(\mc{X}_6))=(99,3)$ and a resolved conifold $\tilde{\mc{X}}_6$ connected by a conifold transition, as the simplest example. Note that $\mc{X}_6$ was used in the construction of flux vacua near the conifold limit in \cite{Demirtas:2020ffz}. The general conifold transition in CY3 geometry has been discussed in \cite{Candelas:1989ug,Candelas:1989js,Candelas:1990pi}, and the physical process has been discussed in type II string compactification~\cite{Strominger:1995cz,Greene:1995hu,Greene:1996dh} and heterotic string compactification~\cite{Anderson:2022bpo} setups.

\begin{figure}
\centering
\includegraphics[height=5.5cm]{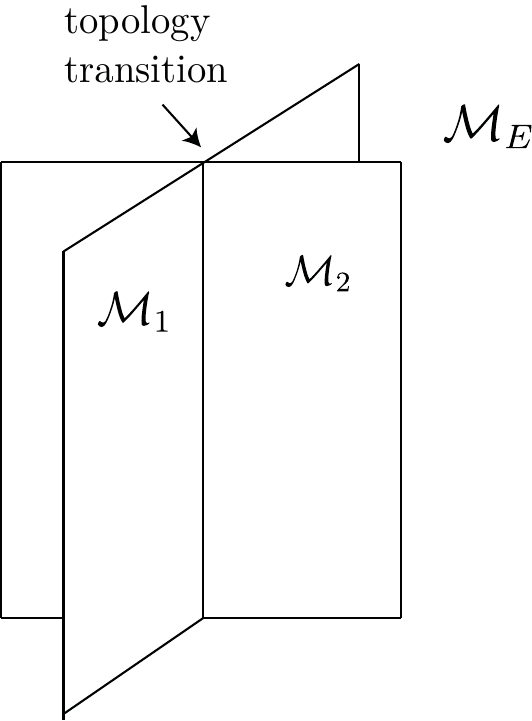}
\caption{A demonstration of the extended moduli space $\mc{M}_E$ containing two subspaces $\mc{M}_1$ and $\mc{M}_2$ that correspond to compact manifolds with different topologies. $\mc{M}_1\cap\mc{M}_2$ corresponds to the locus where the topology transition happens.}\label{f:modulispace}
\end{figure}

We investigate the phase transition in $\mc{M}_E$, in a IIB orientifold compactification setup. Note that we are required to break down to 4d $\mc{N}=1$ SUSY to realize a non-flat scalar potential, otherwise the moduli space is totally flat and no isolated metastable vacuum would exist. In our setup the breaking of SUSY is realized by introducing O7-branes which are away from the conifold point and do not intersect the exceptional cycles. 

Now we illustrate the topological transition more explicitly. We start from $\mc{X}_6$ that is considered as the deformed conifold, and take the conifold limit $\chi_6$. In the process a rigid 3-cycle $\mc{C}$ with the topology of $S^3$ is shrunk to zero volume, and the D3-brane wrapping mode over such an $S^3$ would become massless. We demonstrate that the zero mode from such a D3-brane wrapping $S^3$ corresponds to a hypermultiplet containing two complex scalar fields, in the absence of $H_3$, $F_3$ flux through the $S^3$, using a brane quantization analysis. 

For the particular $\mc{X}_6$ with $(h^{1,1}(\mc{X}_6),h^{2,1}(\mc{X}_6))=(99,3)$, the prepotential $\mc{F}$ of the 4d $\mc{N}=2$ SUGRA can be computed by similar techniques as in \cite{Alvarez-Garcia:2020pxd}, for a point near the conifold and the large complex structure (LCS) point, denoted by $\mc{F}_c$ and $\mc{F}_{\rm LCS}$ respectively. From the form of $\mc{F}_c$ near the conifold, we read off that in $\mc{X}_6$ there are two linearly dependent $S^3$ cycles shrunk to zero volume simultaneously, leading to two hypermultiplets in the 4d $\mc{N}=2$ SUGRA. After SUSY breaking from 4d $\mc{N}=2$ to $\mc{N}=1$ by adding the O7-brane, they are projected to two complex scalar fields $\phi_a$ $(a=1,2)$. The physical importance of $\phi_a$ is that near the conifold limit, these scalar fields are very light, and should be treated on an equal footing as the geometric moduli.

In order to describe how do the brane wrapping modes $\phi_a$ affect $\mc{F}$ and the 4d $\mc{N}=1$ superpotential, we propose the following approximation: we compute the difference $\mc{F}_{\mc{C}}=\mc{F}_{c}-\mc{F}_{\rm LCS}$ across the conifold limit and the LCS point of $\mc{X}_6$, and interpret $\mc{F}_{\mc{C}}$ as the contribution to the prepotential by integrating out the D3-brane wrapping modes over the 3-cycle $\mc{C}$. On a generic point of the CY3 moduli space, we subtract $\mc{F}_{\mc{C}}$ from $\mc{F}$, to define a new prepotential $\tilde{F}=\mc{F}-\mc{F}_{\mc{C}}$ that lacks the correction from brane wrapping modes. Subsequently, we add the contributions of $\phi_a$ 
\be
K_{\mc{C}}\propto(\phi_1^\dagger\phi_1+\phi_2^\dagger\phi_2)\ ,\ W_{\mc{C}}\propto Z\phi_1\phi_2\,
\ee
explicitly into the 4d $\mc{N}=1$ K\"{a}hler potential and superpotential, obtaining $W$ as a holomorphic function of the complex moduli and $\phi_a$.

After such treatment, we compute the $F$-term scalar potential $V_F$ with the no-scale property in  the tree-level   approximation as in the KKLT scenario~\cite{Kachru:2003aw} and its generalization recently~\cite{Demirtas:2019sip,Alvarez-Garcia:2020pxd,Demirtas:2020ffz}. The local minima of $V_F$ can be solved by
\be
D_{\tilde{M}}W=0\,,
\ee
where $\tilde{M}$ ranges over all the complex structure moduli, all the wrapping modes and $\tau$. 

For the choice of flux $(F_3,H_3)$, we require that it does not go through the vanishing $S^3$, because it is not sensible to have a non-vanishing flux through a cycle with zero volume. Under such a choice of flux parameters, denoting the volume of the shrinking 3-cycle by Vol$(S^3)=Z$, qualitatively there are three types of vacuum solutions:

\begin{enumerate}
\item $Z\neq 0$, $\phi_a\neq 0$: this case corresponds to a non-geometric phase, similar to a mixed branch in 4d $\mc{N}=2$ SUSY field theories. We obtain new metastable vacua in this category for the explicit example of $\mc{X}_6$.

\item $Z=0$, $\phi_a\neq 0$: such a case actually corresponds to the geometric phase of the resolved conifold $\tilde{\mc{X}}_6$, where $\langle\phi_1\rangle=\langle\phi_2\rangle$ can be actually interpreted as the new K\"{a}hler moduli parameter in $\tilde{\mc{X}}_6$. In this sense we have finished the conifold transition without explicitly studying the geometry of $\tilde{\mc{X}}_6$.

\item $Z\neq 0$, $\phi_a=0$: this corresponds to the usual geometric phase of the deformed conifold. However, we do not find any local minimum, due to our difference with the literature~\cite{Alvarez-Garcia:2020pxd} that we do not allow flux through the exceptional $S^3$.

\end{enumerate}

Now we ought to discuss the quantum tunneling process between two vacua, e.g. the true vacuum $V_t$ with $\phi_a\neq 0$, $Z\neq 0$ and the false vacuum $V_f$ with $\phi_a\neq 0$, $Z=0$. However, up to this step, all local minima have $V=0$ due to the tree-level approximation. In order to have a non-zero quantum tunneling rate between different vacua, we need to include higher-order corrections to induce a difference between the scalar potential of the two vacua $V_f$ and $V_t$.

We estimate the subleading corrections to the scalar potential from various sources \cite{Gao:2022uop,Junghans:2022exo}, including warping corrections, genuine loop effects that are largely insensitive to the UV completion of the 10d theory, as well as local $\alpha'$ corrections and higher-dimensional operators.  The leading correction is the  standard $\alpha'^3$ contribution arising from the compactification of the higher-curvature term $R^4_{10}$ in the 10d type IIB action \cite{Antoniadis:1997eg,Becker:2002nn,Cicoli:2008va}, which has been taken into account in the whole procedure of moduli stabilization. The next leading correction in our parameter region is the so-called genuine loop correction \cite{Gao:2022uop, Bansal:2024uzr}, which leads to $\delta V \sim -\frac{|W_0|^2}{\mc{V}^{10/3}}$. Given that the vacuum $V_t$ with $\phi_a\neq 0$, $Z\neq 0$ has $|W_0|^2>0$ and the vacuum $V_f$ with $\phi_a\neq 0$, $Z=0$ has $W_0=0$, we find that indeed after the genuine loop correction is taken into account, $V_t<V_f$ as expected, and hence the non-geometric phase is the true vacuum.

Now we use a thin wall approximation to estimate the quantum tunneling rate\cite{Coleman:1977py,Coleman:1980aw,Parke:1982pm}
\begin{eqnarray}
    \Gamma\sim\mc{O}(10^2)e^{-\ms{B}}\,.
\end{eqnarray}
As in our case the gravitational effect is significant, the tunneling between the Minkowski vacuum and the AdS vacuum is prevented without any uplift of the vacuum energy. What's more, after uplift the vacua to the dS vacua, we can not obtain the extreme value of $\ms{B}$ at a stable point with $\frac{\dd \ms{B}}{\dd \rho}=0$, so we can only approximate the upper bound of $\ms{B}$ at the boundary 
\begin{equation}
    \ms{B}_g=2\pi^2\sigma\left(\frac{3}{\kappa_4^2}\right)^{\frac{3}{2}}\delta^{-\frac{3}{2}}+\frac{12\pi^2}{\kappa_4^4}\left[ \frac{\epsilon^{\frac{3}{2}}-\delta^{\frac{3}{2}}}{(\delta-\epsilon)\delta^{\frac{3}{2}}}+\frac{1}{\delta} \right]\,.
\end{equation}
If we assume that the uplift $\delta$ is of the same order as $\epsilon=\delta V=V_f-V_t$, we can obtain the lower bound $\Gamma_g \gtrsim e^{-\mc{O}(10^{21})}$, which leads to a much larger tunneling rate than $\Gamma_n\sim e^{-\mc{O}(10^{33})}$ without any gravitational effect.

Hence in conclusion, in the regime considered in this paper (KKLT scenario), the quantum tunneling process across different phases in $\mc{M}_E$ is heavily suppressed. In order to increase $\delta V$ and get large, observable effects in e.g. primordial gravitational waves~\cite{Roshan:2024qnv}, it is possibly better to work in the Large Volume Scenario(LVS)  \cite{Balasubramanian:2005zx,Conlon:2005ki} with a much larger $|W_0|$, or in a strongly coupled string theory setup. Nonetheless the detailed computations of K\"{a}hler moduli stabilization would be more involved.

The structure of this paper is as following. In section~\ref{sec:flux-comp} we review the setups of type IIB orientifold compactifications to get 4d $\mc{N}=1$ SUGRA flux vacua, where we carefully write out all the coefficients with mass units. We also review the higher-order corrections present in such scenarios. In section~\ref{sec:quan_D3} we derive the BPS D3-brane wrapping modes over the exceptional $S^3$ in the resolved conifold, from the quantization of the DBI action for a general D$p$-brane. In section~\ref{sec:conifold-t} we describe our proposal to quantify the physical procedure of a conifold transition, by subtracting the contributions of the D3-brane wrapping modes from the prepotential, and then add a K\"{a}hler potential $\mc{K}_{\mc{C}}$ and superpotential $W_{\mc{C}}$ back in. In section~\ref{sec:example} we analyze in 
detail for the case of $\mc{X}_6$ with Hodge numbers $(h^{1,1}(\mc{X}_6),h^{2,1}(\mc{X}_6))=(99,3)$. We compute the scalar potential from the superpotential at the tree-level, and present analytical computations of metastable vacua in the different geometric and non-geometric regimes. In section~\ref{sec:quantum-t} we present a simplified analysis of quantum tunneling between different local minima of the scalar potential in the extended moduli space, including the higher-order correction effects. Finally in section~\ref{sec:discussions} we discuss a number of caveats and future directions.

\section{Type IIB flux compactifications}
\label{sec:flux-comp}

\subsection{Type IIB orientifold compactifications}

In this section, we will first briefly review the 4d $\mc{N}=1$ supergravity from flux compactifications of type IIB string theory on orientifolds of Calabi-Yau threefolds \cite{McAllister:2023vgy}. 

In string compactifications, we can obtain the 4d supergravity through the dimension reduction of the 10d supergravity \cite{McAllister:2023vgy}. The tree level 10d IIB supergravity action in the Einstein frame is \cite{Blumenhagen:2013fgp}
\begin{equation}
\begin{aligned}
S_{\mathrm{IIB}}= & \frac{1}{2 \kappa_{10}^2} \int d^{10} x \sqrt{-\op{det}G_E}\left[R-\frac{\partial_M \tau \partial^M \bar{\tau}}{2(\operatorname{Im} \tau)^2}-\frac{1}{2} \frac{\left|G_3\right|^2}{\operatorname{Im} \tau}-\frac{1}{4}\left|F_5\right|^2\right] \\
& +\frac{1}{8 i \kappa_{10}^2} \int \frac{1}{\operatorname{Im} \tau} C_4 \wedge G_3 \wedge \bar{G}_3\,,
\end{aligned}
\end{equation}
Here we have the 10d gravitational coupling constant
\begin{equation}
    \kappa_{10}^2=\frac{l_s^8}{4\pi}\,,
\end{equation}
the axiodilaton
\begin{equation}
    \tau=C_0+i e^{-\Phi}\,,
\end{equation}
and the string coupling constant
\begin{equation}
    g_s=e^{\Phi}\,.
\end{equation}
The relation between the metrics in the string frame and the Einstein frame is
\begin{equation}
    G_{s,IJ}=G_{E,IJ} e^{\Phi/2}\,.
\end{equation}

In the orientifold construction, the 10d IIB supergravity compactifies on $M_4\times \mc{X}_6$, with $M_4$ the 4d Minkowski space and $\mc{X}_6$ the Calabi-Yau 3-fold. The CY3$\mc{X}_6$ admits a holomorphic involution $\sigma:\mc{X}_6\rightarrow \mc{X}_6$, whose fixed loci are points and/or divisors in $\mc{X}_6$. Then the massless fields in this construction should be invariant under 
\begin{equation}
    \mc{O}=\sigma\Omega_{\op{ws}}(-1)^{F_L}\,.
    \label{orientifold_proj}
\end{equation}
Here $\Omega_{\op{ws}}$ is the reversal of orientation on the string worldsheet. $F_L$ is the left-moving fermion number operator. Such an action of $\mc{O}$ on Dolbeault cohomology classes defines even and odd eigenspaces $H_{\pm}^{p,q}$ with dimensions $h_{\pm}^{p,q}$. We list the massless fields shown in Table \ref{tab:orienti_multi}. For simplicity, we can choose the orientifold action with $h_-^{1,1}=h_+^{2,1}=0$, and all the massless fields are the geometric moduli and the axiodilaton $\tau$.
\begin{table}
    \centering
    \begin{tabular}{c|c}
        Field & Number of fields\\\hline
        K\"ahler moduli $T$ & $h_+^{1,1}$ \\
        complex structure moduli $U$ & $h_-^{2,1}$\\
        two forms $B$ & $h_-^{1,1}$\\
        vector multiplets $V$ & $h_+^{2,1}$\\
        axiodilaton $\tau$ &$1$
    \end{tabular}
    \caption{$\mc{N}=1$ multiplets in O3/O7 orientifolds}
    \label{tab:orienti_multi}
\end{table}

\subsection{Scalar potential}

The action of 4d $\mc{N}=1$ supergravity is highly constrained by supersymmetry, which reads
\begin{equation}
    S_{4d}=\frac{1}{\kappa_4^2}\int \frac{1}{2}R *1 - \mc{K}_{M\bar{N}}D\phi^M\wedge*D\phi^{\bar{N}}-\frac{1}{2}\op{Re}f_{\tilde{M}\tilde{N}} F^{\tilde{M}}\wedge *F^{\tilde{N}}-\frac{1}{2}\op{Im}f_{\tilde{M}\tilde{N}} F^{\tilde{M}}\wedge F^{\tilde{N}}-(V_F+V_D) *1\,.
    \label{4d_n=1_sugra}
\end{equation}
Here $M,N$ range over all chiral multiplets and $\tilde{M},\tilde{N}$ range over all vector multiplets.  This action can be completely determined by the holomorphic superpotential $W$, the real K\"ahler potential $\mc{K}$ and the holomorphic gauge coupling $f_{\tilde{M}\tilde{N}}$ \cite{Freedman:2012zz}. The scalar potential includes the F-term potential and the D-term potential
\begin{equation}
    V_F\propto e^{\mc{K}}\left(\mc{K}^{M\bar{N}} D_MW D_{\bar{N}}\bar{W}-3|W|^2\right)\,,\;D_M W=W_M+W\mc{K}_M\,,\;\d_{M}=\frac{\d}{\d \phi_{M}}\,,
    \label{F_term_sugra}
\end{equation}
\begin{equation}
    V_D\propto \frac{1}{2}\left(\op{Re}f\right)^{-1\,\tilde{M}\tilde{N}}\mc{P}_{\tilde{M}}\mc{P}_{\tilde{N}}\,,\;\mc{P}_{\tilde{M}}=(\mc{K}_M+\frac{W_M}{W})(T_{\tilde{M}})^{MN}\phi_N=\frac{1}{W}D_MW (T_{\tilde{M}})^{MN}\phi_N\,,
    \label{D_term_sugra}
\end{equation}
\begin{equation}
    W_M=\d_MW\,,\;\mc{K}_M=\d_M\mc{K}\,,\;\mc{K}_{MN}=\d_M\d_{\bar{N}}\mc{K}\,,\;\mc{K}^{MN}=(\mc{K}_{M\bar{N}})^{-1}\,.
\end{equation}
Here $T_{\tilde{M}}$ denote the non-abelian gauge symmetry generators.

After dimension reduction, we can find that at the tree level, the 4d Planck mass is
\begin{equation}
    M_p^2=\frac{1}{\kappa_4^2}=\frac{4\pi\mc{V}}{l_s^2}\,.
\end{equation}
Here $\mc{V}$ is the volume of $\mc{X}_6$. From Dirac quantization and supersymmetry \cite{Becker:2006dvp,Blumenhagen:2013fgp}, we have the flux quantization condition
\begin{equation}
    \frac{1}{l_s^2}\int H_3\in H^3(\mc{X}_6,\mb{Z})\,,\quad\frac{1}{l_s^2}\int F_3\in H^3(\mc{X}_6,\mb{Z})\,.
\end{equation}

The concrete form of the above massless scalars in Table \ref{tab:orienti_multi} is shown in \cite{Candelas:1990pi,Becker:2006dvp}. In the CY3-fold $\mc{X}_6$, up to an overall factor, there is a unique holomorphic $(3,0)$ form $\Omega\in H^{3,0}(\mc{X}_6)$. We take $\alpha_{\tilde{i}}$, $\beta^{\tilde{j}}$, $\tilde{i},\tilde{j}=0,\dots h^{2,1}$ as a symplectic basis of $H^3(\mc{X}_6,\mb{Z})$, which satisfies $\int\alpha_{\tilde{i}}\wedge\beta^{\tilde{j}}=-\delta_{\tilde{i}}^{\tilde{j}}$, such that for any 3-forms $\Pi_1,\Pi_2$, 
\begin{equation}
    \int \Pi_1\wedge\Pi_2=\left(\int\Pi_1\wedge\alpha_{\tilde{i}},\int\Pi_1\wedge\beta^{\tilde{j}}\right)\left(\begin{array}{cc}
        0 & -I \\
        I & 0
    \end{array}\right)\left(\int\Pi_2\wedge\alpha_{\tilde{i}},\int\Pi_2\wedge\beta^{\tilde{j}}\right)^T=\Pi_1\cdot\Sigma\cdot\Pi_2\,.
\end{equation}
The intersection matrix is denoted as
\begin{equation}
    \Sigma=\left(\begin{array}{cc}
        0 & -I \\
        I & 0
    \end{array}\right)\,.
\end{equation}
And with abuse of notation, we use $\Pi$ to denote the 3-form $\Pi$ and as well as the vector $\left(\int\Pi\wedge\alpha_{\tilde{i}},\int\Pi\wedge\beta^{\tilde{j}}\right)$ in the basis $\alpha_{\tilde{i}}$, $\beta^{\tilde{j}}$. In this form, by choosing the overall factor s.t. $\int \Omega\wedge\alpha_0=1$, with $\mc{N}=2$ supersymmetry, we have
\begin{equation}
    \Omega=\left(\int\Omega\wedge\alpha_{\tilde{i}},\int\Omega\wedge\beta^{\tilde{j}}\right)=\left(1,U^i,2\mc{F}-U^i\d_{U^i}\mc{F},\d_{U^i}\mc{F}\right)\,,\;i=1\dots h^{2,1}\,.
\end{equation}
calculated with Picard-Fuchs equations \cite{Hosono:1993qy,Hosono:1994av,Hosono:1994ax}. Here $\mc{F}$ is the 4d $\mc{N}=2$ prepotential in the form of
\begin{equation}
    \mc{F}(U)=\mathcal{F}_{0}(U)+\mathcal{F}_{\text {inst }}(U)\,
\end{equation}
with
\begin{equation}
    \mathcal{F}_{0}(U)=-\frac{1}{3!} K_{ijk} U^i U^j U^k+\frac{1}{2} a_{ij} U^i U^j+ b_i U^i+\frac{\zeta(3) \chi(\mc{Y}_6)}{2(2 \pi i)^3}\,,
    \label{F0_general}
\end{equation}
Here $K_{ijk}$ denotes the triple intersection numbers of the mirror Calabi-Yau threefold $\mc{Y}_6$, while 
\begin{equation}
    a_{ij}=\frac{1}{2}\left\{
    \begin{aligned}
        &K_{iij}\,,\;i\geq j\\
        &K_{ijj}\,,\;i<j\\
    \end{aligned}
    \right.\,,\;\;\;b_i=\frac{1}{24}\int_{\mc{Y}_6}c_2(\mc{Y}_6)\wedge \tilde{\beta}_i \,.
\end{equation}
\begin{equation}
    \mathcal{F}_{\text {inst }}(U)=-\frac{1}{(2 \pi i)^3} \sum_{\mathbf{q} \in \mathcal{M}(\mc{Y}_6)} n^0_{\mathbf{q}} \operatorname{Li}_3\left(e^{2 \pi i \mathbf{q}\cdot \mathbf{U}}\right)\,.
    \label{F_inst_general}
\end{equation}
The coefficients $n^0_{\mbf{q}}$ are the genus-zero Gopakumar-Vafa (GV) invariants \cite{Gopakumar:1998ii,Gopakumar:1998jq} of $\mc{Y}_6$.

With nontrivial flux $F_3, H_3$, we can obtain the Gukov-Vafa-Witten superpotential \cite{Gukov:1999ya}
\begin{equation}
    W_{\op{flux}}=\frac{1}{l_s^2}\int G_3\wedge\Omega\,,
\end{equation}
\begin{equation}
    G_3=F_3-\tau H_3\,.
\end{equation}
There are also quantum corrections from Euclidean D$p$-branes and gaugino condensation to the superpotential\cite{McAllister:2023vgy}. But these corrections are not explicitly considered in this work, as we only try to stabilize the brane wrapping modes and complex structure moduli. At our level of approximation we have
\begin{equation}
    W\approx W_{\op{flux}}\,.
    \label{W_approx}
\end{equation}

The tree level K\"ahler potential is
\begin{equation}
    \mc{K}_{0}=-2\log \mc{V}(T,\bar{T})-\log \mc{V}_{\Omega}(U,\bar{U})-\log(-i(\tau-\bar{\tau}))\,.
\end{equation}
Here 
\begin{equation}
    \mc{V}=\mc{V}_0+\delta \mc{V} \,,\quad \mc{V}_0=\frac{1}{3!}\int_{\mc{X}_6}J\wedge J\wedge J\,,\quad \delta \mc{V}=-\frac{\zeta(3)\chi(\mc{X}_6)}{4(2\pi)^3g_s^{3/2}}+\text{higher order terms}\,.
\end{equation}
$\mc{V}_0$ is the volume of the CY3-fold $\mc{X}_6$ and can be determined by the K\"ahler form $J$. $\delta \mc{V}$ is the quantum correction to this volume.
\begin{equation}
    \mc{V}_{\Omega}=\int i\Omega\wedge\bar{\Omega}=i\Omega\cdot\Sigma\cdot\bar{\Omega}
\end{equation}
is the volume of the mirror CY3-fold $\mc{Y}_6$ together with the quantum correction.

Ignoring the higher order corrections to the K\"ahler potential, we use the approximation \cite{Demirtas:2019sip,Alvarez-Garcia:2020pxd,Demirtas:2020ffz}
\begin{equation}
    \mc{K}\approx \mc{K}_{0}\,.
    \label{K_approx}
\end{equation}
Then the leading term of the F-term potential in the current convention\footnote{In \cite{Conlon:2005ki},  the refined Einstein frame $G_{s,IJ}=G_{e,IJ}e^{(\Phi-\Phi_0)/2}$ is used. The translation between the Einstein frame and the refined Einstein frame is shown in Appendix \ref{sec:F_potential}.} reads
\begin{equation}
    V_F=\frac{1}{4\pi\kappa_4^2}e^{\mc{K}}\left(\mc{K}^{M\bar{N}} D_MW D_{\bar{N}}\bar{W}-3|W|^2\right)\,,
\end{equation}
with $M,N$ labeling all complex scalars. At the tree level, the K\"ahler potential has the no-scale property \cite{Cremmer:1983bf}
\begin{equation}    \mc{K}_{0}^{a\bar{b}}\mc{K}_{0,a}\mc{K}_{0,\bar{b}}=3\,,
    \label{no_scale_prop}
\end{equation}
with $a,b$ ranging over all K\"ahler moduli $T$. So with the above approximations (\ref{W_approx},\ref{K_approx}), we have
\begin{equation}
    V_F=\frac{1}{4\pi\kappa_4^2}e^{\mc{K}}\left(\mc{K}^{\tilde{M}\bar{\tilde{N}}} D_{\tilde{M}}W D_{\bar{\tilde{N}}}\bar{W}\right)\,.
    \label{F_poten_orienti}
\end{equation}
Here $\tilde{M},\tilde{N}$ range over all complex scalars other than the K\"ahler moduli.

\subsection{Higher order corrections}
\label{subsec:higherorder}

There are several types of corrections when considering type IIB orientifold compactifications. For simplicity, we  follow \cite{Gao:2022uop} to present the results in the context of the Large Volume Scenario (LVS)  \cite{Balasubramanian:2005zx,Conlon:2005ki} although the results can be applied to more general situations.   

First, there are warping corrections due to the classical  backreaction of fluxes and branes to the background geometry.  Warping corrections to the K\"ahler potential take the form of a series of terms suppressed by $1/\lambda$, $1/\lambda^2$ etc., where $\lambda$ is the volume of a generic 4-cycle. The complete series  is expected to not affect the scalar potential since warping corrections respect the no-scale structure \cite{Giddings:2001yu}.  However,  such corrections do put a parametric constraint on the compact geometry when one considers inflationary scenarios with anti-D3 brane uplift \cite{Junghans:2022exo,Gao:2022fdi,Bansal:2024uzr}.

Second, there are genuine loop corrections. From the 10d perspective, these arise from string loops or from fields localized on branes propagating through the compactified space. From the 4d viewpoint, they correspond to the effects of integrating out the tower of KK modes. A key distinguishing feature of these corrections is their non-locality in the higher-dimensional theory: they cannot be captured by local operators in 10d or on a brane, and are analogous to Casimir energy effects.  Such genuine one-loop effects correct the K\"ahler potential as
$
   \delta \mc{K}_{\rm loop} \sim  \frac{1}{\sqrt{\lambda} \mc{V}} \sim \mathcal{O}(\frac{1}{\mc{V}^{4/3}})\,.
$
The genuine one-loop correction agrees with the so-called winding type of the Berg-Haack-Pajer (BHP) conjecture for loop corrections \cite{Berg:2005ja,Berg:2007wt,Cicoli:2007xp,Cicoli:2008va,Cicoli:2009zh}, although it is independent of the specific brane setup assumed in their analysis. In the Einstein frame, this will result in a correction to the scalar potential up to a $g_s/8\pi$ factor as:
\begin{equation}
\label{deltaV-loop}
    \delta V_{\rm loop} \sim  -\frac{ |W_0|^2}{ 4 \pi \kappa^2_4 \mc{V}^2} \frac{1}{\sqrt{\lambda} \mc{V}} \sim  - \frac{|W_0|^2}{4 \pi \kappa^2_4} \mc{O}(\frac{1}{\mc{V}^{10/3}})\,.
\end{equation}
This correction provides the main contribution when evaluating the quantum tunneling rate between different vacua, as discussed in Section \ref{sec:quantum-t}.

Third, there are  $\alpha'$ corrections, which originate  either from higher-dimensional local operators in the 10d bulk, on D-branes, O-planes, and at their intersection loci, or  from the high-momentum regions of loop integrals, appearing as counterterms required to  renormalize ultraviolet divergences.  Notably, there is no sharp distinction between  local $\alpha'$ corrections that are part of the classical action and those arising as loop-induced counterterms. It thus appears natural to group all such corrections, whether fundamental or loop-generated, that can be associated with higher-dimensional local operators under  the name `local $\alpha'$ corrections'.

Independent of genuine loop effects, higher-curvature terms also correct the 4d Einstein–Hilbert term obtained after compactification. In particular, the dimensional reduction of the  $R^4_{10}$ term in the higher-curvature part of the 10d Type IIB action yields the  well-known string tree-level BBHL correction, as well as its one-loop counterpart \cite{Antoniadis:1997eg,Becker:2002nn}:
\begin{equation}
\left(\frac{M^{2}_{10}}{g_s^{3/2}}+M_{10}^2g_s^{1/2}\right)R_{\text{external}} \int \dd^6 x R_{\text{internal}}^{3}\sim \left(\frac{M^{2}_{10}}{g_s^{3/2}}+M_{10}^2g_s^{1/2}\right) R_{\text{external}}\,.
\end{equation}
 These corrections exhibit the same scaling behavior as the corresponding contributions to the K\"ahler potential, which arise after  performing Weyl rescaling to the 4d Einstein frame. In this way, we obtain \cite{Becker:2002nn, Cicoli:2008va}
\begin{eqnarray}
    \delta V^{R^4_{10}}_{\rm local} & \sim & \frac{|W_0|^2 }{4 \pi \kappa^2_4}(g_s^{-3/2}+g_s^{1/2})\mc{O}(\frac{1 }{\mc{V}^3})\,.
    \label{deltaV-BBHL}
\end{eqnarray}
The first term of (\ref{deltaV-BBHL}) is also known as the standard $\alpha'$ correction \cite{Balasubramanian:2005zx,Conlon:2005ki, Cicoli:2008va}, which  takes the form 
\begin{equation}
\label{deltaV-leading}
    \delta V_{\alpha'} \simeq \frac{1}{4 \pi \kappa^2_4}\frac{3 \xi |W_0|^2}{4 g_s^{3/2} \mc{V}^3}\,,
\end{equation}
with $\xi = - \frac{\chi(\mc{X}_6)\zeta(3)}{2(2\pi)^3}$. Here the Riemann zeta function is $\zeta(3) \sim 1.2$ and $\chi(\mc{X}_6)$ is the Euler number of the Calabi-Yau $\mc{X}_6$. Note that this term is absorbed into the quantum corrected volume $\mc{V}$ of $\mc{X}_6$ later.

Similarly, higher-curvature terms of the form $R_{p+1}^n$ localized  on the $p$-brane should  also be considered, as their dimensional reduction contributes to the coefficient of the 4d Einstein-Hilbert term $R_4$.  The worldvolume action of a $p$-brane contains two types of curvature corrections:  those coming from  the curvature corrections of the Dirac–Born–Infeld (DBI) action  \cite{Bachas:1999um}, and  topological curvature terms in the Wess-Zumino (WZ) action \cite{Bachas:1999um,Green:1996dd}.  The latter vanish in our setup, as we assume there is  no flux through the brane as discussed in the following section \ref{sec:quan_D3}. 
In our case, we consider D7-branes wrapping a  4-cycle with a typical length scale $L$, which is at most of order $\mc{O}(\mc{V}^{1/6})$. The leading  operator $R_8^2$  on the D7-brane yields corrections to the K\"ahler potential $\delta \mc{K}^{R^2_8}_{\rm local} \sim g_s \mc{O}(\mc{V}^{-2/3})$, which scale similarly to the  KK-type corrections proposed in  the BHP conjecture:
\begin{equation}
  \delta V^{R^2_8}_{\rm local} \sim  \frac{g_s^2 |W_0|^2}{4 \pi \kappa^2_4} \mc{O}( \frac{1}{\mc{V}^{10/3}} )\,.
\end{equation}
Additionally,  there is  a marginal operator  $R_8^4$ can be  generated on  the D7-brane,  leading to a logarithmic enhancement of the form $\ln(M_{10}g_s^{1/4} \mc{V}^{1/6})$ in the coefficient of $R_4$  \cite{Gao:2022uop}. This operator introduces logarithmic corrections to  both the K\"ahler potential and the scalar potential 
\begin{equation}
\label{deltaV-marginal}
 \delta V^{R^4_8}_{\rm local} = \frac{|W_0|^2}{4 \pi \kappa^2_4}  \ln(M_{10}g_s^{1/4} \mc{V}^{1/6}) \mc{O}(\frac{1}{\mc{V}^{10/3}})\,.
\end{equation}
 It is important to note that this logarithm  term arises at the same order in $\alpha'$ as genuine one-loop effects, i.e., $\alpha'^4$.   A similar analysis can be extended to D-brane/O-plane intersection systems.

In this paper, we incorporate the $\alpha'$ correction (\ref{deltaV-leading}) in the process of moduli stabilization to obtain the vacuum. 
The next-to-leading corrections arise from various sources and can be comparable in magnitude. In certain regions of parameter space, 
the appearance of a logarithmic enhancement in (\ref{deltaV-marginal}), associated to the marginal operator in the D-brane/O-plane system, would be dominant compared to genuine loop effects. However, after moduli stabilization calculated in Section (\ref{subsec:moduli_stabilize}),
 the leading correction to this vacuum arises from the genuine loop correction $\delta V_{\rm{loop}}$ (\ref{deltaV-loop}) when considering the quantum tunneling between different vacua.

\section{The D$p$-brane wrapping modes}
\label{sec:quan_D3}

It is known that we cannot fully quantize the $p$-branes $(p\geq 2)$ due to the nonlinearity of the equation of motion~\cite{Taylor:2001vb}. However, if we only consider the massless spectrum and ignore the higher order corrections, we can quantize the zero modes of the $p$-branes wrapping $p$-cycles $\mc{C}$ in $\mc{X}_6$ with a semi-classical approximation \cite{Shifman:2012zz}. This basic idea is briefly shown in \cite{Witten:1996qb}, and in this section we will derive the details for later reference. 

The worldvolume theory of D$p$-branes consists of the Dirac-Born-Infeld (DBI) action and the Chern-Simons (CS) action~\cite{Blumenhagen:2013fgp}. In the string frame, the bosonic part of the DBI action is 
\begin{equation}
    S_{\mathrm{DBI}}=-T_{\op{Dp}} \int_{\mathscr{W}} d^{p+1} \sigma e^{-\Phi(X)} \sqrt{-\operatorname{det}\left(g_{\mu\nu}(X)+2 \pi \alpha^{\prime} \mathscr{F}_{\mu\nu}(X)\right)}\,,
    \label{DBI_action_string_frame}
\end{equation}
with 
\begin{equation}
    \begin{aligned}
        &g_{\mu\nu}=\partial_\mu X^I \partial_\nu X^J G_{s,IJ}\,, \quad b_{\mu \nu}=\partial_\mu X^I \partial_\nu X^J B_{IJ}\,,\\
        &2\pi \alpha' \mathscr{F}=b+2\pi \alpha' F,\quad T_{\op{Dp}}=\frac{2\pi}{l_s^{p+1}}\,,\\
        &\mu,\nu=0,\dots p\,,\quad I,J=0,\dots 9\,.
    \end{aligned}
\end{equation}
 Here $X^I(\sigma^{\mu})$ is the embedding of the world-volume $\mathscr{W}(\sigma^{\mu})$ into the whole spacetime $M_4(x^0,x^u)\times \mc{X}_6(x^m)\,, u=1,\dots 3\,,\; m=1,\dots 6$. $F=\dd A$ is the field strength of the vector field $A$ on the world-volume. The CS action is 
\begin{equation}
    S_{\mathrm{CS}}=\left.T_{\op{Dp}} \int_{\mathscr{W}} \operatorname{tr}\left(e^{2 \pi \alpha^{\prime} \mathscr{F}}\right) \wedge \sqrt{\frac{\hat{A}\left(4 \pi^2 \alpha^{\prime} R_T\right)}{\hat{A}\left(4 \pi^2 \alpha^{\prime} R_N\right)}} \wedge \bigoplus_q C_q\right|_{p+1}\,.
\end{equation}
When $|2\pi\alpha'\mathscr{F}|\ll 1$, we can expand
\begin{equation}
\begin{aligned}
S_{\mathrm{CS}}= & T_{\op{Dp}} \int_{\mathscr{W}}\left( C_{p+1}+2 \pi \alpha^{\prime} C_{p-1} \wedge \operatorname{tr} \mathscr{F}\right. \\
& \left.+\frac{1}{2}\left(2 \pi \alpha^{\prime}\right)^2 C_{p-3} \wedge\left[\operatorname{tr} \mathscr{F} \wedge \mathscr{F}+\frac{1}{48}\left(\operatorname{tr} R_T \wedge R_T-\operatorname{tr} R_N \wedge R_N\right)\right]+\ldots\right)\,.
\end{aligned}
\end{equation}
Here $R_T,R_N$ are the tangent and normal bundles of $\mathscr{W}\hookrightarrow M_4\times \mc{X}_6$. 

\subsection{First quantization with semi-classical approximation}

In this work, we are only interested in the case where the D$p$-branes wrap the p-cycle in the compact space $\mc{X}_6$, i.e. the world-volume $\mathscr{W}$ of the D$p$-branes is $\mc{C\times T}$. $\mc{C}$ is the $p$-cycle in $\mc{X}_6$ and $\mc{T}$ is the time axis in $M_4$.

For simplicity, we take the simplification that all the closed string fields, $G$, $B$, $\Phi$ and $C_p$ are treated as background fields, and their dynamics are ignored. In addition, we only consider the case that there is no flux through $\mc{C}$. We freeze $F=0$, $B=0$, such that $\mathscr{F}=0$. We only consider the BPS D$p$-brane wrapping over the  $p$-cycle that is a special Lagrangian manifold \cite{Marino:1999af}\footnote{The proof of this condition on 3-cycles is reviewed in Appendix \ref{sec:BPS_3-cycle}.}. We will quantize the zero modes in the vicinity of this BPS configuration $X^I=x^I(\sigma^{\mu})$\footnote{In general cases, we also need to consider the zero modes of the 1-form field $\ms{F}$. But as we only consider D3-branes wrapping $S^3$, there is no such zero mode. So to show the relation between the moduli space and the spectrum, we ignore $\ms{F}$ in the following derivation.}
\begin{equation}
    \begin{aligned}
        X^{u}&=x^u=\op{const}\,,\; x^0=\sigma^0\,.\\
        X^m=&X_0^m+\delta X^m=X_0^{m}(\sigma^{\hat{\mu}})+ c_{\mc{A}}\zeta_{\mc{A}}^m(\sigma^{\hat{\mu}})\,,\;\hat{\mu}=1,\dots p\,,\\
        \mathscr{F}&=0\,,\;\Phi=\langle\Phi\rangle\,.
    \end{aligned}
\end{equation}
Here $X_0^{m}(\sigma^{\hat{\mu}})$ is the configuration of a $p$-dimensional special Lagrangian submanifold $\mc{C}$ in $\mc{X}_6$. $\zeta_{\mc{A}}$ is the small variation which keeps the special Lagrangian condition. In other words, $c_{\mc{A}}$ are the coordinates of the moduli space of the special Lagrangian submanifold $\mc{C}$ in $\mc{X}_6$. The dimension of the moduli space can be determined by the first Betti number of the special Lagrangian submanifold $\op{dim}\mc{M}_{\mc{C}}=b_1(\mc{C})$ \cite{ellingsrud2002calabi}. In our work, we only focus on the trivial case $\mc{C}\simeq S^3$ with $\op{dim}\mc{M}_{\mc{C}}=b_1(S^3)=0$. But in principle, the quantization is general while the case with a nontrivial moduli space is more  difficult. In summary, the zero modes of the D$p$-brane on $\mc{C\times T}$ are composed of $x^u,c_{\mc{A}}$.

In the full quantization, we need to consider the more general variations $\delta X^u\neq 0$ and $\delta X^{m}$ changing the special Lagrangian condition, but such variations do not correspond to zero modes and can be massive even when the volume $\mc{V}(\mc{C})\rightarrow 0$. So we could expect that such modes will generate heavy fields in the effective theory and can be integrated out. 


To do the quantization explicitly, we should first transform the string frame action into the Einstein frame action. As mentioned above $G_s=G_E e^{\Phi/2}$, we can rewrite the action in the form of $g_{\mu\nu}=g_{E,\mu\nu} e^{\Phi/2}$
\begin{equation}
    S_{\op{DBI}}=-\mu_{\op{Dp}}\int_{\ms{W}}\dd^{p+1}\sigma \sqrt{-\op{det}(g_E)}\,,
\end{equation}
\begin{equation}
    S_{\op{CS}}=\mu_{\op{Dp}}e^{-\frac{p-3}{4}\Phi}\int_{\ms{W}}C_{p+1}\,.
\end{equation}
Here we denote the brane tension in the Einstein frame $\mu_{\op{Dp}}=T_{\op{Dp}}e^{\frac{p-3}{4}\Phi}$.

As we need a 4d effective theory in the end, we will integrate out the degrees of freedom in $\mc{X}_6$. Such degrees of freedom are related to $\sigma^{\hat{\mu}}$ in the worldvolume. In the semi-classical approximation, to quantize the effective theory, we only consider the variation $\delta X^I$ by setting the zero modes dynamical $x^u=x^u(\sigma^0)$, $c_{\mc{A}}=c_{\mc{A}}(\sigma^0)$. Then the induced metric on the worldvolume can be expanded as
\begin{equation}
    \begin{aligned}
        g_{E,00}&=-1+G_{E,uv}\delta \dot{X}^u \delta \dot{X}^v+G_{E,mn} \delta\dot{X}^m \delta \dot{X}^n\,,\\
        g_{E,\hat{\mu}0}&=G_{E,mn}\frac{\d \delta X^m}{\d \sigma^{\hat{\mu}}}\delta \dot{X}^n\,,\\
        g_{E,\hat{\mu}\hat{\nu}}&=g_{E,0,\hat{\mu}\hat{\nu}}+G_{E,mn}\frac{\d \delta X^m}{\d \sigma^{\hat{\mu}}}\frac{\d \delta X^n}{\d \sigma^{\hat{\nu}}}\,,\;\hat{\mu},\hat{\nu}=1,\dots3\,.
    \end{aligned}
\end{equation}
We can expand the determinant of the matrix
\begin{equation}
\operatorname{det}(1+M)=1+\operatorname{tr}(M)+\frac{1}{2}(\operatorname{tr} M)^2-\frac{1}{2} \operatorname{tr}\left(M^2\right)+\cdots
\end{equation}
So up to the leading order
\begin{equation}
    \begin{aligned}
        &\sqrt{-\op{det}\left(g_{E,\hat{\mu}\hat{\nu}}\right)}=\sqrt{-\op{det}\left(g_{E,0,\hat{\mu}\hat{\nu}}\right)}\cdot\sqrt{\op{det}\left(1+g_{E,0}^{\hat{\mu}\hat{\rho}}\delta g_{E,\hat{\rho}\hat{\nu}}\right)}\\
        &=\sqrt{-\op{det}\left(g_{E,0,\hat{\mu}\hat{\nu}}\right)}\left(1+\frac{1}{2}g_{E,0}^{\hat{\mu}\hat{\nu}}\delta g_{E,\hat{\mu}\hat{\nu}}+\mc{O}(\left|\delta g_E\right|^2)\right)\\
        &=\sqrt{-\op{det}\left(g_{E,0,\hat{\mu}\hat{\nu}}\right)}\left[1+\frac{1}{2}\left(
        -G_{E,uv}\delta \dot{X}^u \delta \dot{X}^v-G_{E,mn} \delta\dot{X}^m \delta \dot{X}^n+g_{E,0}^{\hat{\mu}\hat{\nu}}G_{E,mn}\frac{\d \delta X^m}{\d \sigma^{\hat{\mu}}}\frac{\d \delta X^n}{\d \sigma^{\hat{\nu}}}
        \right)+\mc{O}(\left|\delta g_E\right|^2)\right]\\
        &=\sqrt{-\op{det}\left(g_{E,0,\hat{\mu}\hat{\nu}}\right)}\left[1+\frac{1}{2}\left(
        -G_{E,uv} \dot{x}^u \dot{x}^v-G_{E,mn} \zeta^m_{\mc{A}} \zeta^n_{\mc{B}} \dot{c}^{\mc{A}}\dot{c}^{\mc{B}}+g_{E,0}^{\hat{\mu}\hat{\nu}}G_{E,mn}\frac{\d \delta X^m}{\d \sigma^{\hat{\mu}}}\frac{\d \delta X^n}{\d \sigma^{\hat{\nu}}}
        \right)+\mc{O}(\left|\delta g_E\right|^2)\right]\,.
    \end{aligned}
\end{equation}

Then we need to integrate over the worldvolume $\mc{C}$. From the definition of the moduli space, when all the time derivatives $\dot{x},\dot{c}$ are zero, the equation of motion $\delta S=0$ should always hold. So we have
\begin{equation}
    \int_{\ms{W}} \dd^{p+1}\sigma \sqrt{-\op{det}\left(g_{E,0,\hat{\mu}\hat{\nu}}\right)} g_{E,0}^{\hat{\mu}\hat{\nu}}G_{E,mn}\frac{\d \delta X^m}{\d \sigma^{\hat{\mu}}}\frac{\d \delta X^n}{\d \sigma^{\hat{\nu}}}=0\,.
\end{equation}
Then we can find that the DBI action becomes
\begin{equation}
    S_{\op{DBI}}\rightarrow-\int \dd \sigma^0\left[ m_{\mc{C}}\left(1-\frac{1}{2}h_{uv}\dot{x}^u\dot{x}^v\right)-\frac{1}{2}h_{\mc{AB}}\dot{c}^{\mc{A}}\dot{c}^{\mc{B}}+\dots\right]\,.
    \label{QM_DBI_boson_zero}
\end{equation}
\begin{equation}
    m_{\mc{C}}=\mu_{\op{Dp}}\mc{V}(\mc{C})\,.
\end{equation}
Here the metric of the moduli space is
\begin{equation}
    h_{uv}=\delta_{uv},\quad h_{\mc{AB}}=m_{\mc{C}}\int_{\mc{C}} \dd^p\sigma \, G_{E,mn} \zeta^m_{\mc{A}}\zeta^n_{\mc{B}}\,.
    \label{wrap_moduli_metric}
\end{equation}
We can notice that $h_{uv}$ is constant, and $x^u$ represent the location in the lower dimensional space. $h_{\mc{AB}}=h_{\mc{AB}}(c)$ depends on the moduli of the wrapping modes. 

As for the CS action, from our setup, it becomes
\begin{equation}
    S_{\op{CS}}=\mu_{\op{Dp}}e^{-\frac{p-3}{4}\Phi}\int_{\ms{W}} C_{p+1}\,.
\end{equation}
From Kaluza-Klein reduction, we know the massless modes are
\begin{equation}
    C_{p+1}=l_s^{p+1}A^{\tilde{i}}\wedge \alpha_{\tilde{i}}\,.
\end{equation}
Here $\alpha_{\tilde{i}}$ range over the basis of $H^p(\mc{X}_6,\mb{Z})$. The intersection number
\begin{equation}
    q_{\mc{C},\tilde{i}}=\int_{\mc{C}}\alpha_{\tilde{i}}\,,
\end{equation}
determines the charge of the wrapping modes under the $\op{U}(1)$ gauge vector $A^{\tilde{i}}_{\mu}$. Choosing the Coulomb gauge $A^{\tilde{i}}_{0}=0$,  the CS action becomes
\begin{equation}
    S_{\op{CS}}\rightarrow \mu_{\op{Dp}}e^{-\frac{p-3}{4}\Phi}l_s^{p+1}\int \dd \sigma^0 q_{\mc{C},\tilde{i}}A^{\tilde{i}}_{u}\dot{x}^u\,.
\end{equation}

In summary, we have obtained the bosonic part of the effective quantum mechanics of the wrapping modes, which is a nonlinear sigma model
\begin{equation}
    \begin{aligned}
    \label{SCQM}
        S_{\mc{C},QM}&=S_{\mc{C},p}+S_{\mc{C},\op{spec}}\,,\\
        S_{\mc{C},p}&=-\int \dd \sigma^0 \left[m_{\mc{C}}\left(1-\frac{1}{2}h_{uv}\dot{x}^u\dot{x}^v\right)-g_{\op{Dp}}q_{\mc{C},\tilde{i}}A^{\tilde{i}}_{u}\dot{x}^u\right]\,,\\
        S_{\mc{C},\op{spec}}&=\int \dd \sigma^0 \frac{1}{2}h_{\mc{AB}}\dot{c}^{\mc{A}}\dot{c}^{\mc{B}}+\dots
    \end{aligned}
\end{equation}
Here $S_{\mc{C},p}$ is the worldline action of a non-relativistic charged particle under a $\op{U}(1)_{\tilde{i}}$ gauge field with the gauge coupling constant\begin{equation}
    g_{\op{Dp}}=e^{-\frac{p-3}{4}\Phi}\mu_{\op{Dp}}l_s^{p+1}\,.
\end{equation}
$S_{\mc{C},p}$ has the generalized momentum
\begin{equation}
    P_u=m_{\mc{C}}\dot{x}_u+g_{\op{Dp}}q_{\mc{C},\tilde{i}}A^{\tilde{i}}_{u}\,.
\end{equation}
Thus we obtain the first quantized Hamiltonian 
\begin{equation}
    H_p=m_{\mc{C}}+\frac{(P_u-g_{\op{Dp}}q_{\mc{C},\tilde{i}}A^{\tilde{i}}_{u})^2}{2m_{\mc{C}}}+\dots\,.
    \label{QM_DBI_1st_quan_H}
\end{equation}
When we take into account of the fermionic part, similar to \cite{Tong:sqm,Dedushenko:2014nya}, there are four fermion zero modes $\psi_1,\psi_2,\psi_1^{\dagger},\psi_2^{\dagger}$ from the broken half of the eight supercharges in the type IIB compactification. Then through a standard argument on supermultiplets \cite{Terning:2006bq}, when the moduli space is trivial, the vacuum state $|0\rangle$ of (\ref{QM_DBI_1st_quan_H}) expands into a hypermultiplet with four states $\{|0\rangle,\psi_1^{\dagger}|0\rangle,\psi_2^{\dagger}|0\rangle,\psi_1^{\dagger}\psi_2^{\dagger}|0\rangle\}$. 

$S_{\mc{C},\op{spec}}$ is the non-linear sigma model on the moduli space $\mc{M_{C}}$. By including the four broken supercharges $Q,\bar{Q},Q^{\dagger},\bar{Q}^{\dagger}$ \cite{Witten:1996qb,Tong:sqm,Hori:2003ic} with the representation $\d,\bar{\d},\d^{\dagger},\bar{\d}^{\dagger}$ in $\mc{M_C}$. As $H_{\mc{C},\op{spec}}=\{\bar{Q},\bar{Q}^{\dagger}\}=\Delta_{\bar{\d}}$, the ground states of the supersymmetric quantum mechanics $S_{\mc{C},\op{spec}}$ are the harmonic forms on $\mc{M_{C}}$. In this way, the moduli space $\mc{M}_{\mc{C}}$ can determine the supermultiplet representations of the wrapping modes. For the deformed conifold, we only consider the case of a rigid $\mc{C}$, i.e. $\mc{M_{C}}=\{\text{Single point}\}$. In this case, there is only one hypermultiplet and $H_{\mc{C}}=H_{p}$.

In our work, we mainly focus on the orientifold compactification, which is 4d $\mc{N}=1$ instead of 4d $\mc{N}=2$. This means that the orientifold action will project out half of the eight supercharges. Then we can expect that half of the four broken supercharges will be projected into $\psi,\psi^{\dagger}$. As a consequence, the hypermultiplet becomes a chiral multiplet $\{|0\rangle,\psi^{\dagger}|0\rangle\}$. Nonetheless, the action of the complex scalars is invariant under the orientifold action, which is still $S_{\mc{C},p}$ (\ref{SCQM}).

\subsection{Second quantization}
\label{sec:second-q}

In the above discussion, with the semi-classical approximation, we obtain the quantum mechanics of the wrapping modes. However, to do moduli stabilization, we need to rewrite these modes as fields in the 4d quantum field theory. This can be realized through a second quantization analysis.

From the Hamiltonian $H_p$, ignoring the higher order corrections, it can be seen that the wave function should satisfy the non-relativistic equation of motion 
\begin{equation}
    i\d_0 \phi=\left(m_{\mc{C}}+\frac{(-i\d_u-g_{\op{Dp}}q_{\mc{C},\tilde{i}}A^{\tilde{i}}_{u})^2}{2m_{\mc{C}}}\right)\phi\,,
\end{equation}
which implies the relativistic version of the equation of motion
\begin{equation}
    \label{2nd-q-eom}(D_{\mu}D^{\mu}+m_{\mc{C}}^2)\phi(x^{\mu})=0\,,\;D_{\mu}=i\d_{\mu}+g_{\op{Dp}}q_{\mc{C},\tilde{i}}A^{\tilde{i}}_{\mu}\,.
\end{equation}
Then we should transform $\phi$ from a function to an operator, following the standard canonical quantization
\begin{equation}
    \phi(x^{\mu})=\sum_{p} a_p\phi_p+b_p^{\dagger}\bar{\phi}_p\,,
\end{equation}
$a_p^{\dagger},b_p^{\dagger}$ are the creation operators of states with momentum $p$.
The effective action should take the form of a massive charged complex scalar, reproducing the equation of motion (\ref{2nd-q-eom})
\begin{equation}
    \begin{aligned}
        S_{\mc{C},p}=&-\int \dd^4 x \phi^{\dagger}(x^{\mu})(D_{\mu}D^{\mu}+m_{\mc{C}}^2)\phi(x^{\mu})\\
        =&-\int \dd^4 x\left(\overline{D^{\mu}\phi}D_{\mu}\phi+m_{\mc{C}}^2\phi^{\dagger}\phi\right)\,.
    \end{aligned}
    \label{EFT_wrapping}
\end{equation}
When there is a nontrivial moduli space $\mc{M}_{\mc{C}}$, we can also perform the second quantization with the equation of motion and effective action
\begin{equation}
    \left[ (D_{\mu}D^{\mu}+m_{\mc{C}}^2)+\Delta_{\bar{\d}} \right]\phi(x^{\mu},c^{\mc{A}})=0\,,
\end{equation}
\begin{equation}
    S_{\mc{C}}=-\int \dd^4 x\dd c \,\phi(x^{\mu},c^{\mc{A}})^{\dagger}\left[ (D_{\mu}D^{\mu}+m_{\mc{C}}^2)+\Delta_{\bar{\d}} \right]\phi(x^{\mu},c^{\mc{A}})\,.
\end{equation}

\section{Conifold transitions in IIB}
\label{sec:conifold-t}

\subsection{Conifold transitions in CY3-folds}
\label{sec:conifold-CY3}

\begin{figure}
    \centering
    \includegraphics[width=1\linewidth]{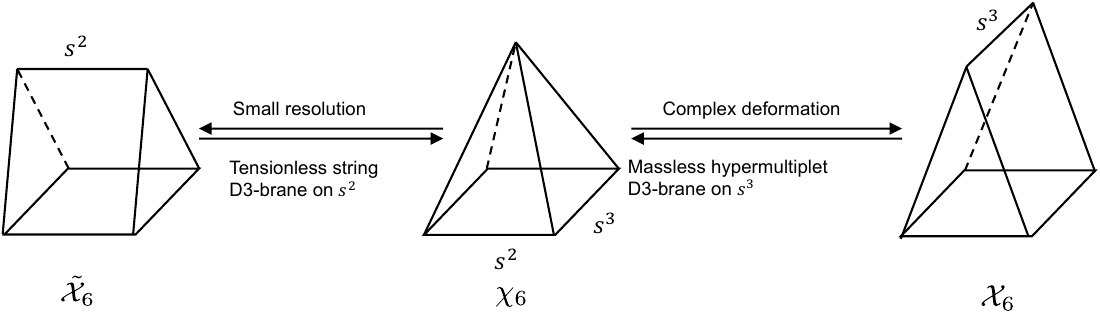}
    \caption{Local geometry of the CY3-fold near the conifold singularity}
    \label{fig:conifold_geometry}
\end{figure}

Conifold transitions between smooth CY3-folds are discussed in \cite{Candelas:1989ug, Candelas:1989js}. The local geometry of the conifold is a real cone $\triangleleft\left(S^2 \times S^3\right)$. As shown in Figure \ref{fig:conifold_geometry}, the compact space $\chi_6$ at the conifold singularity is singular. It can be made smooth by either the complex deformation, in which the exceptional 3-cycle $S^3$ has non-zero volume or the small resolution, in which the exceptional 2-cycle $S^2$ has non-zero volume. In these ways we obtain two different smooth CY3-folds $\mc{X}_6$ and $\tilde{\mc{X}}_6$. 

When $\mc{X}_6$ reaches the singularity $\chi_6$, a number of 3-cycles are shrunk to points, and the number of the complex moduli would be reduced. Then as we resolve $\chi_6$ to $\tilde{\mc{X}}_6$, a number of new 2-cycles are generated, and the number of K\"ahler moduli will be increased. In summary, when we start from $\mc{X}_6$ and reach $\tilde{\mc{X}}_6$, $\Delta h^{2,1}<0$ and $\Delta h^{1,1}>0$. Thus  $\mc{X}_6$ and $\tilde{\mc{X}}_6$ are topologically inequivalent. 


In general cases, from $\mc{X}_6$ to $\chi_6$, there is a series of 3-cycles $\mc{C}_a$, $a=1,\dots, n_{\mc{C}}$\footnote{As there are too many different indices in this paper, without ambiguity, in the alphabet there are four kinds of different indices: $1. a,b,c\dots$ $2. i,j,k\dots$ $3. m,n,p\dots$ $4. u,v,w\dots$} shrinking to points. These 3-cycles are not linearly  independent, as there is a set of nontrivial linear relations $L_{\alpha a}\mc{C}_a=0$, $\alpha=1,\dots, n_L$, $a=1,\dots, n_{\mc{C}}$. After the conifold transition, $n_L$ additional linear independent 2-cycles will be created. Hence in this process, we have $\Delta h^{2,1}=-(n_{\mc{C}}-n_L)$, $\Delta h^{1,1}=n_L$.

\subsection{Conifold transitions in 4d $\mc{N}=2$}
\label{subsec:conifold_4dn=2}

The conifold transitions in the 4d $\mc{N}=2$ case, i.e. IIB on a CY3-fold, have been well studied~\cite{Strominger:1995cz,Greene:1995hu,Greene:1996dh}. In our work, we mainly focus on the complex deformation of the conifold.

The volume of the 3-cycle is proportional to the complex structure moduli. When we go from $\mc{X}_6$ to $\chi_6$, some of the complex structure moduli $Z_a\rightarrow 0$, $a=1,\dots, n_{\mc{C}}$. As the corresponding 3-cycles are linearly dependent, we can expect that these moduli have the same linear relations $L_{\alpha a}Z_a=0$, $\alpha=1,\dots, n_L$, $a=1,\dots, n_{\mc{C}}$. In other words, these moduli can be represented by a series of linearly independent basis $Z_a=N_{a\hat{a}}Z_{\hat{a}}$, $\hat{a}=1,\dots,n_{\mc{C}}-n_L$. In this limit, the prepotential is singular and in the form of 
\begin{equation}
    \mc{F}\sim \sum_{a=1}^{n_{\mc{C}}}\frac{1}{4\pi i} Z^2_a\log(2\pi i Z_a)+\text{holomorphic terms}\,.
\end{equation}
Such a result can be observed through the general form (\ref{F_inst_general}) as discussed in \cite{Demirtas:2020ffz}. In the cases where there is only one $Z_{\hat{a}}=Z$ and the GV invariants of the corresponding cycle $\mc{C}$ satisfy $n^0_{n\mc{C}}=0$ when $n>n_0$, then as $Z\rightarrow 0$, only finitely many terms in (\ref{F_inst_general}) become non-negligible. Then from Euler's reflection formula
\begin{equation}
-\frac{\mathrm{Li}_3\left(e^{2 \pi i Z}\right)}{(2 \pi i)^3}=\frac{Z^2}{4 \pi i} \ln (-2 \pi i Z)-\frac{1}{(2 \pi i)^3} \sum_{n=0}^{\infty} \frac{\hat{\zeta}(3-n)}{n!}(2 \pi i Z)^n\,,
\end{equation}
we obtain 
\begin{equation}
    \mc{F}\overset{Z\rightarrow 0}{\sim} \frac{n_{\mc{C}}Z^2}{4\pi i}\log (2\pi iZ)+\text{holomorphic terms}\,.
\end{equation}
Here $n_{\mc{C}}$ can be obtained from the instanton computation and can be understood as the number of vanishing $\mc{C}$ in the $Z\rightarrow 0$ limit.

Similar to the argument in \cite{Seiberg:1994rs}, these singularities in $\mc{F}$ imply that there are some additional massless fields at this point in the moduli space, and the effective field theory description is no longer valid. 
So if we want to discuss the physics near conifold singularities and the conifold transition process, we need a new effective field theory description with these additional massless fields. As argued in \cite{Strominger:1995cz}, on the complex deformation side, the additional massless fields are the hypermultiplets from D3-branes wrapping the vanishing 3-cycles $\mc{C}_a$, whose masses are $m_{a}=T_{\op{D3}}\mc{V}(\mc{C}_a)\propto|Z_a|$ and vanish at the singularity $Z_a=0$.

Now we describe the realization of the conifold transition in this new effective theory~\cite{Greene:1995hu}. For simplicity, near the conifold singularity, we only consider the relevant vector multiplets and the hypermultiplets from D3-brane wrapping modes. Through the quantization of the zero modes of D3-branes reviewed in Section~\ref{sec:quan_D3}, we find that a D3-brane wrapping on a single $S^3$ leads to one new hypermultiplet. Hence in the new effective theory, there are $n_{\mc{C}}$ new hypermultiplets. At the point of  $Z_{\hat{a}}=0$, $\phi_{a}^{\zeta}=0$, we obtain the effective action of $n_{\mc{C}}$ massless hypermultiplets and $n_{\mc{C}}-n_L$ massless vector multiplets. 


From the above effective action (\ref{EFT_wrapping}), with the form of 4d $\mc{N}=2$ SUSY \cite{Tachikawa:2013kta}, we can expect that the superpotential and the K\"ahler potential of the hypermultiplets are
\begin{equation}
    W_{\mc{C}}\sim \sum_{a} Z_a\phi_a^{1}\phi_a^2=\sum_{a,\hat{a}}N_{a\hat{a}}Z_{\hat{a}}\phi_a^{1}\phi_a^2\,,\;\mc{K}_{\mc{C}}\sim \sum_{a\hat{a}}\left(\phi_a^{1\dagger}e^{N_{a\hat{a}}V_{\hat{a}}}\phi_a^{1}+\phi_a^{2\dagger}e^{-N_{a\hat{a}}V_{\hat{a}}}\phi_a^{2}\right)\,.
\end{equation}
In rigid 4d $\mc{N}=2$ SUSY, the potential reads~\cite{Freedman:2012zz}
\begin{equation}
    \begin{aligned}
    \label{4dN2VF}
        V_F&\sim \sum_{\hat{a}}\left|\d_{Z_{\hat{a}}}W_{\mc{C}}\right|^2+\sum_{a}\left|\d_{\phi_a^{1}}W_{\mc{C}}\right|^2+\sum_{a}\left|\d_{\phi_a^{2}}W_{\mc{C}}\right|^2\\
        &\sim\sum_{\hat{a}}|\sum_a N_{a\hat{a}}\phi_a^{1}\phi_a^2|^2+ \sum_a|\sum_{\hat{a}}N_{a\hat{a}}Z_{\hat{a}}\phi_a^{1}|^2+\sum_a|\sum_{\hat{a}}N_{a\hat{a}}Z_{\hat{a}}\phi_a^2|^2\,,
    \end{aligned}
\end{equation}
\begin{equation}
    \begin{aligned}
    \label{4dN2VD}
        V_D&\sim \sum_{\hat{a}\hat{b}}(\op{Re}f^{-1})^{\hat{a}\hat{b}} \left[\sum_aN_{a\hat{a}}(\mc{K}_{\mc{C},\phi_a^{1}}\phi_a^{1}-\mc{K}_{\mc{C},\phi_a^{2}}\phi_a^{2})\right]\left[\sum_aN_{a\hat{b}}(\mc{K}_{\mc{C},\phi_a^{1}}\phi_a^{1}-\mc{K}_{\mc{C},\phi_a^{2}}\phi_a^{2})\right]\\
        &\sim\sum_{\hat{a}\hat{b}}(\op{Re}f^{-1})^{\hat{a}\hat{b}} \left[\sum_aN_{a\hat{a}}(\phi_a^{1\dagger}\phi_a^{1}-\phi_a^{2\dagger}\phi_a^{2})\right]\left[\sum_aN_{a\hat{b}}(\phi_a^{1\dagger}\phi_a^{1}-\phi_a^{2\dagger}\phi_a^{2})\right]\,.
    \end{aligned}
\end{equation}

Similar to the Coulomb branch and Higgs branch classification in 4d $\mc{N}=2$ SUSY field theory~\cite{Tachikawa:2013kta}, there are two simple classes of supersymmetric vacua $Z$: $\phi_a^{\zeta}=0, Z_{\hat{a}}\neq 0$ and $\phi_a^{\zeta}\neq 0,Z_{\hat{a}}=0$. 

In the first case, $Z_{\hat{a}}$s are massless and $\phi_a^\zeta$s have the masses shown in (\ref{4dN2VF}).  

In the second case, 
we have $2(n_{\mc{C}}-n_{L})$ real constraints from $V_F$, $n_{\mc{C}}-n_{L}$ real constraints from $V_D$ and $n_{\mc{C}}-n_{L}$ real constraints from the gauge invariance. 
Consequently there are only $4n_L$ real flat directions, corresponding to $n_L$ massless hypermultiplets. The remaining $n_{\mc{C}}-n_L$ massless hypermultiplets will combine with the $n_{\mc{C}}-n_L$ massless vector multiplets $Z_{\hat{a}}$ into $n_{\mc{C}}-n_L$ massive vector multiplets through the Higgs mechanism~\cite{Strominger:1995cz}. Hence there are $n_L$ additional massless hypermultiplets and $n_{\mc{C}}-n_L$ fewer massless vector multiplets in the new string vacua with $\phi_{a}^{\zeta}\neq 0$, $Z_{\hat{a}}=0$. 

In type IIB CY3 compactifications, there are $h^{2,1}$ vector multiplets and $h^{1,1}+1$ hypermultiplets. So we can expect that when we transit from the vacua $\phi_{a}^{\zeta}=0$, $Z_{\hat{a}}\neq 0$ to the new vacua $\phi_{a}^{\zeta}\neq 0$, $Z_{\hat{a}}=0$, the shift of the Hodge numbers is $\Delta h^{2,1}=-(n_{\mc{C}}-n_L)$, $\Delta h^{1,1}=n_L$. This exactly matches the physical descriptions of conifold transitions above. Hence we can expect that there are two branches of vacua in this new effective theory, i.e. the branch $\phi_{a}^{\zeta}= 0$, $Z_{\hat{a}}\neq 0$ corresponds to the complex deformation side $\mc{X}_6$, and the branch $\phi_{a}^{\zeta}\neq 0$, $Z_{\hat{a}}=0$ corresponds to the small resolution side $\tilde{\mc{X}}_6$. 

\begin{figure}
    \centering
    \includegraphics[width=0.5\linewidth]{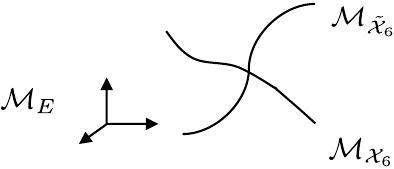}
    \caption{The string vacua $\mc{M}_{\mc{X}_6}$, $\mc{M}_{\tilde{\mc{X}}_6}$ with different topologies in the extended space $\mc{M}_{E}$}
    \label{fig:extended}
\end{figure}

Thus in order to describe the transition between different phases in $\mc{M}_E$, we need to take into account additional massive fields. The enlarged field space is no longer the moduli space of CY3-folds, which is flat at the tree level, but a larger extended space $\mc{M}_E$ with a non-flat potential. Then by solving the equation $V=0$, we can obtain a set of different vacua, which correspond to string compactifications with different topologies or  non-geometric phases. In the above example, as shown in Figure~\ref{fig:extended}, the extended space includes the CY3 moduli of $X$ and the $n_{\mc{C}}$ additional massless hypermultiplets from D3-brane wrapping modes. The vacua in $\mc{M}_E$ include the moduli space $\mc{M}_{\mc{X}_6}=\left\{\phi_{a}^{\zeta}= 0, Z_{\hat{a}}\neq 0\right\}$ of $\mc{X}_6$ and the moduli space $\mc{M}_{\tilde{\mc{X}}_6}=\left\{\phi_{a}^{\zeta}\neq 0, Z_{\hat{a}}= 0\right\}$ of $\tilde{\mc{X}}_6$.

When the quantum corrections are included, it is possible that the vacua in $\mc{M}_E$ are not only $\mc{M}_{\mc{X}_6}$ and $\mc{M}_{\tilde{\mc{X}}_6}$. Just like the 4d $\mc{N}=2$ SUSY field theory may have the mixed branch vacua, there may be some vacua $\mc{M}_{\op{non}}\subset \left\{\phi_{a}^{\zeta}\neq 0, Z_{\hat{a}}\neq 0\right\}$. Such vacua $\mc{M}_{non}$ do not have a clear geometric interpretation, and hence are interpreted as non-geometric string vacua.

\subsection{Conifold transition in 4d $\mc{N}=1$}
\label{subsec:conifold_4dn=1}

After the orientifold action is taken into consideration, the effective action is transformed to a 4d $\mc{N}=1$ one. The most significant difference is that a lot of complex scalars are projected out during the process. K\"ahler moduli and complex structure moduli are all embedded in chiral multiplets instead of hypermultiplets and vector multiplets. However, as the orientifold projection (\ref{orientifold_proj}) is only defined on the string worldsheet, we do not explicitly derive the D3-brane wrapping modes in the orientifold compactifications. 

However, if we assume that the 4d $\mc{N}=2$ description of the conifold transition is still valid, as the D3-brane wrapping modes can be transformed into K\"ahler moduli in the 4d $\mc{N}=2$ scenario, one hypermultiplet from the D3-brane wrapping modes should also be projected into one chiral multiplet analogous to the K\"ahler moduli. There can be argued in two ways. First, in the 4d $\mc{N}=2$ case, the extra massless D3-brane wrapping modes in $\mc{X}_6$ become new K\"ahler moduli in $\tilde{\mc{X}}_6$. In 4d $\mc{N}=2$ both the wrapping modes and K\"ahler moduli are hypermultiplets. However, in the 4d $\mc{N}=1$ case, the K\"ahler moduli are projected into the chiral multiplets, so the wrapping modes should also be projected into chiral multiplets. Second, in the 4d $\mc{N}=2$ case, some hypermultiplets from wrapping modes combine with the vector multiplets from complex structure moduli to form massive vector multiplets. But in the 4d $\mc{N}=1$ case, the complex structure moduli are projected into chiral multiplets, hence we only need chiral multiplets to realize the Higgs mechanism which combines the massless chiral multiplets and massless vector multiplets. In conclusion, the 4d $\mc{N}=2$ D3-brane wrapping modes should be projected into chiral multiplets in the 4d $\mc{N}=1$ case.

Then we can make a guess of the possible form of the superpotential and K\"ahler potential in the 4d $\mc{N}=1$ case. For the example in \ref{subsec:geometry_and_scalar_potential}, the flux and the orientifold do not intersect the vanishing $S^3$, and all complex structure moduli are projected into chiral multiplets. So we can expect that the $\mc{N}=1$ correction is subdominant in this topological transition. Such an argument is not quantitative, and it is possible that some new phenomenon beyond this work can take place \cite{Kaufmann:2026fli,Kaufmann:2026mha}. We will discuss the effect of the $\mc{N}=1$ correction in future work.

With the above argument, we can assume that the effective action of the D3-brane (\ref{EFT_wrapping}) is still valid at least at the tree level. Then the two complex scalars in a hypermultiplet can be combined into one complex scalar $(\phi_a^1,\phi_a^2)\rightarrow \phi_a=c_1 \phi_a^1+c_2\phi_a^2$, $|c_1|^2+|c_2|^2=1$. According to the semi-classical approximation in sec \ref{sec:second-q}, the leading order of the superpotential and K\"ahler potential of the wrapping modes will be schematically
\begin{equation}
    W_{\mc{C}}\sim \sum_{a} Z_a\phi_a\phi_a=\sum_{a,\hat{a}}N_{a\hat{a}}Z_{\hat{a}}\phi_a\phi_a\,,\;\mc{K}_{\mc{C}}\sim \sum_{a}\phi_a^{\dagger}\phi_a\,.
    \label{Wrapping_poten_general}
\end{equation}
And the scalar potential is 
\begin{equation}
    \begin{aligned}
        V_F&\sim \sum_{\hat{a}}\left|\d_{Z_{\hat{a}}}W_{\mc{C}}\right|^2+\sum_{a}\left|\d_{\phi_a}W_{\mc{C}}\right|^2\\
        &=\sum_{\hat{a}}|\sum_aN_{a\hat{a}}\phi_a\phi_a|^2+ \sum_a|\sum_{\hat{a}}N_{a\hat{a}}Z_{\hat{a}}\phi_a|^2\,.
    \end{aligned}
\end{equation}
The D-term potential vanishes because there is no relevant vector field in our setup. With a similar argument to the 4d $\mc{N}=2$ case, there are two branches of vacua, $\phi_a=0$, $Z_{\hat{a}}\neq 0$ and $\phi_a\neq0$, $Z_{\hat{a}}= 0$. The first one is $\mc{X}_6$ and the second one is $\tilde{\mc{X}}_6$. When we take into account of the higher order corrections and gravity effect, there will be vacua with $\phi_a\neq 0,Z_{\hat{a}}\neq 0$, which are in a non-geometric phase. We will show such non-geometric vacua in Section \ref{subsec:moduli_stabilize}.

In the 4d $\mc{N}=2$ case, we can only obtain $\Delta h^{1,1}$ and $\Delta h^{2,1}$ from the arguments in Section \ref{sec:conifold-CY3}. In the 4d $\mc{N}=1$ case, the details of the orientifold projection also affect the physical theory, and there exists an ambiguity about how the shift $\Delta h^{1,1}$ decomposes into $\Delta h^{1,1}=\Delta h^{1,1}_++\Delta h^{1,1}_-$. Such a decomposition may depend on how we choose the orientifold which is not the topic of our work.

\subsection{D3-brane wrapping modes in 4d $\mc{N}=1$ supergravity}
\label{sec:D3-wrap-4dN1}

Until now, we have not discussed the effects of the complex structure moduli $Z_{\hat{a}}$ and the supergravity. In later computations, we find that even though we do not include the higher order quantum corrections, the tree level supergravity action is enough to stabilize most of the moduli.
In this section we consider the inclusion of the D3-brane wrapping mode into the 4d $\mc{N}=1$ supergravity, starting with the exact form of the superpotential and K\"ahler potential. For simplicity, we only consider the case where there is only a single $Z_{\hat{a}}=Z$, which is enough for our example and is the common case in the literature, see e.g. \cite{Demirtas:2020ffz}.

In section \ref{sec:second-q}, we have obtained the action of the complex scalars from D3-brane wrapping modes 
\begin{equation}
    S_{\mc{C}}=-\sum_a\int D\phi_a\wedge *D\phi_a^{\dagger}+m_{\op{D3}}^2|\phi_a|^2*1\,.
\end{equation}
The mass of the complex scalars is
\begin{equation}
    m_{\op{D3}}=\mu_{\op{D3}}\mc{V}(\mc{C})l_s^3\,,
\end{equation}
where $\mc{V}(\mc{C})$ is the volume of the BPS 3-cycle $\mc{C}$
\begin{equation}
    \mc{V}(\mc{C})=\sqrt{\frac{8\mc{V}}{\mc{V}_{\Omega}}}\left|\int_{\mc{C}}\Omega\right|=\sqrt{\frac{8\mc{V}}{\mc{V}_{\Omega}}}\left|Z\right|\,.
    \label{3-cylce_volume_corr}
\end{equation}
The derivation is shown in \cite{Becker:1995kb} and will be reviewed in Appendix \ref{sec:BPS_3-cycle}. 
Here, the volume of the 3-cycle receives a correction due to the modification of the classical Calabi–Yau volume $\mc{V}_0$ of $\mc{X}_6$ into the volume $\mc{V}$ with quantum corrections. This correction is also necessary to ensure consistency with the quantum correction of   $\mc{V}_{\Omega}$, which arises from  the last term of the prepotential (\ref{F0_general}) \cite{Hosono:1994av}.

When the geometry is $M_4\times \mc{X}_6$, i.e. $W=0$ and $\phi_a=0$, there are no further corrections to the above mentioned action. In this case, the K\"ahler potential should be
\begin{equation}
    \frac{1}{\kappa_4^2}\d_{\phi_a}\overline{\d_{\phi_b}}\mc{K}_C=\delta_{ab}\,,\;\mc{K}_{\mc{C}}=\kappa_4^2(\sum_a\phi_a\phi_a^{\dagger})\,.
\end{equation}

The mass term $m_{\op{D3}}^2|\phi|^2$ can be generated from the F-term potential at the tree level
\begin{equation}
    \frac{1}{4\pi\kappa_4^4}e^{\mc{K}}\mc{K}_{\mc{C}}^{a\bar{b}} D_{a}W_{\mc{C}} D_{\bar{b}}\bar{W}_{\mc{C}}\overset{\phi_a\rightarrow 0}{\sim}m_{\op{D3}}^2(|\phi_1|^2+|\phi_2|^2)\,.
\end{equation}
at the tree level,
\begin{equation}
    e^{-\mc{K}}\overset{\phi_a\rightarrow 0}{\sim}2\op{Im}\tau\mc{V}^2\mc{V}_{\Omega}\,.
\end{equation}
So we can obtain
\begin{equation}
    \mu_{\op{D3}}^2 l_s^6\frac{8\mc{V}}{\mc{V}_{\Omega}}|Z|^2(\sum_a|\phi_a|^2) e^{-\mc{K}}4\pi \kappa_4^6\subset \d_{\phi_{a}} W_{\mc{C}}\d_{\phi_{b}^{\dagger}}\bar{W}_{\mc{C}}\,.
\end{equation}
We can assume that the superpotential should have a similar form to the above 4d $\mc{N}=1$ SUSY, i.e. 
\begin{equation}
    W_{\mc{C}}\propto Z(\sum_a \phi_a^2)\,,
\end{equation}
which implies
\begin{equation}
\label{ZphiinWc}
    e^{i\theta}T_{D3}l_s^3\sqrt{32\pi i\kappa_4^6\mc{V}^3\op{Im}\tau} \frac{1}{2}Z(\sum_a \phi_a^2)\subset W_{\mc{C}}\,.
\end{equation}
Here the phase factor $e^{i\theta}$ comes from replacing the absolute value with the complex number $|Z|\rightarrow Z$, which can be absorbed by a redefinition  $\phi_a\rightarrow e^{i\theta/2}\phi_a$. In the above expression, we only use $\subset$ instead of $=$ because that the left hand side of (\ref{ZphiinWc}) involves $\op{Im}\tau$, which is not a holomorphic function. To arrive at a holomorphic superpotential, we assume that one can replace $i\op{Im}\tau\rightarrow\tau$ to obtain the full superpotential. This operator will generate new terms in the potential which vanish when $C_0=0$. Then we can obtain the superpotential 
\begin{equation}
    W_{\mc{C}}=l_s^2\sqrt{\frac{\tau}{2}} Z(\sum_a\phi_a^2)+\text{higher order terms}\,.
\end{equation}
In our explicit example in the next section, we are discussing the simplest case of $n_{\mc{C}}=2$, and the superpotential can be rewrite in a more compact form by a redefinition of $\phi_a$
\begin{equation}
    W_{\mc{C}}=l_s^2\sqrt{2\tau} Z\phi_1\phi_2+\text{higher order terms}\,.
\end{equation}

\section{Example}
\label{sec:example}

In this section, we will give an example of moduli stabilization with the D3-brane wrapping modes taken into account. For simplicity, we choose the example in \cite{Demirtas:2020ffz}, which is reviewed briefly in Appendix~\ref{sec:99-3}. And we use a moduli stabilization process similar to that in \cite{Demirtas:2019sip,Alvarez-Garcia:2020pxd}, where there is non-trivial flux through the vanishing cycle $\mc{C}$ and the D3-brane wrapping modes are integrated out. In our construction, we instead consider the case with no flux through $\mc{C}$ and not integrating out wrapping modes. In the following calculations, we will set the string length $l_s=2\pi\sqrt{\alpha'}=1$. 

\subsection{Geometry and scalar potential}
\label{subsec:geometry_and_scalar_potential}


On the Calabi-Yau threefold $\mc{X}_6$ with $(h^{1,1}(\mc{X}_6),h^{2,1}(\mc{X}_6))=(99,3)$\cite{Demirtas:2020ffz}, there exists a choice of the O7-brane, s.t. $h^{1,1}_{+}=h^{1,1}$, $h^{2,1}_{-}=h^{2,1}$, and the D3-brane tadpole bound is $Q_{\op{D3}}=52$.
\begin{equation}
    \begin{aligned}
        &K_{333}=K_{233}=K_{223}=4,\;K_{133}=K_{123}=K_{122}=2,\;K_{222}=3,\\
        &a_{33}=a_{23}=2\,,\;a_{12}=a_{13}=1\,,\;a_{22}=\frac{3}{2}\,,\\
        &b=\left(1,\frac{7}{4},\frac{13}{4}\right),\;\xi=-\frac{12 i \zeta(3)}{ \pi^3}\,.
    \end{aligned}
\end{equation}
From \cite{Demirtas:2020ffz}, we can find that the first few genus zero GV invariants are 
\begin{equation}
    n^0_{(1,0,0)}=2\,,n^0_{(0,1,0)}=252\,,n^0_{(0,0,1)}=2\,.
\end{equation}
For the other genus zero GV invariants, $n^0_{(0,0,k)}=0$ for $k\geq 2$, which comes from the fact that $Z=0$ is the point of conifold singularity.
Near the conifold singularity,
\begin{equation}
    \mc{F}=\mc{F}_0+\mc{F}_{\op{inst}}\,,
\end{equation}
\begin{equation}
\mathcal{F}_0=-\frac{1}{3!}K_{ijk}U^i U^jU^k+\frac{1}{2}a_{ij}U^iU^j+b_i U^i+\xi\,.
\label{F0_eg}
\end{equation}
\begin{equation}
    \begin{aligned}
        \mc{F}_{\op{inst}}&=\mc{F}_{\mc{C}}+\mc{F}_1+\text{higher order terms}\\
        &=-\frac{1}{(2\pi i)^3}\left[ 2\op{Li}_3(e^{2\pi i U^1})+252\op{Li}_3(e^{2\pi i U^2})+2\op{Li}_3(e^{2\pi i Z}) \right]+\text{higher order terms}\,,
    \end{aligned}
    \label{F_inst_eg}
\end{equation}
\begin{equation}
    \mc{F}_{\mc{C}}=-\frac{1}{(2\pi i)^3} 2\op{Li}_3(e^{2\pi i Z})=2\left[\frac{Z^2}{4\pi i }\log(-2\pi i Z)-\frac{1}{(2\pi i)^3}\sum_{n=0}^{\infty}\frac{\hat{\zeta}(3-n)}{n!}(2\pi i Z)^n\right]\,.
\end{equation}
Here $\hat{\zeta}(k):=\zeta(k)$ when $k\neq 1$ and $\hat{\zeta}(1)=\frac{3}{2}$.
\begin{equation}
    \mc{F}_1= -\frac{1}{(2\pi i)^3}\left[ 2\op{Li}_3(e^{2\pi i U^1})+252\op{Li}_3(e^{2\pi i U^2}) \right]=-\frac{i}{4\pi^3}e^{2\pi i U^1}-\frac{63 i}{2 \pi^3}e^{2\pi i U^2}+\mc{O}(e^{4\pi iU})\,.
\end{equation}
If we expect that $|U^1|\sim |U^2|$, then in the $\mc{O}(e^{4\pi iU})$ term, the leading term corresponds to $n^0_{(0,2,0)}=-9252$, 
\begin{equation}
    \mc{F}_{1,\op{high}}=\frac{2313i}{2\pi^3}e^{4\pi i U^2}+\cdots
\end{equation}
From the genus zero GV invariants, we can find that there are $n_{\mc{C}}=n^0_{(0,0,1)}=2$ wrapping modes. From the discussions in Section \ref{sec:D3-wrap-4dN1}, the corresponding K\"ahler potential and superpotential are
\begin{equation}
    \mc{K}_{\mc{C}}=\kappa_4^2(\phi_1^{\dagger}\phi_1+\phi_2^{\dagger}\phi_2)\,,
    \label{kahler_wrap_eg}
\end{equation}
\begin{equation}
    W_{\mc{C}}=\sqrt{2\tau}Z\phi_1\phi_2\,.
    \label{superpoten_wrap_eg}
\end{equation}
This form is equivalent to the above form (\ref{Wrapping_poten_general}) up to linear recombination of $\phi_a$.

As we choose to not integrate out the wrapping modes and keeping the dependence on $\phi_a$, we should remove $\mc{F}_{\mc{C}}$ from $\mc{F}$ the prepotential, and obtain a new prepotential $\tilde{\mc{F}}$
\begin{equation}
    \tilde{\mc{F}}=\mc{F}_0+\mc{F}_{1}\,,\quad \mc{F}=\mc{F}_0+\mc{F}_{\mc{C}}+\mc{F}_{1}\,,
\end{equation}
For the 4d $\mc{N}=2$ case, it is well known that $\mc{F_C}$ comes from integrating out $\mc{K}_{\mc{C}}$ and $W_\mc{C}$ \cite{Hori:2003ic,Hattab:2024ssg}. And as there are no other possible quantum corrections near the conifold singularity, we can expect that $\mc{F_C}$ still comes from integrating out $\mc{K_C}, W_\mc{C}$ after projecting into the 4d $\mc{N}=1$ theory. From the prepotential, we define the following holomorphic 3-forms
\begin{equation}
    \begin{aligned}
        \Omega&=\left(1,U^1,U^2,Z,2\mc{F}-U^i\d_{U^i}\mc{F},\d_{U^1}\mc{F},\d_{U^2} \mc{F},\d_{Z} \mc{F}\right)\,,\\
        \Omega_0&=\left(1,U^1,U^2,Z,2\mc{F}_0-U^i\d_{U^i}\mc{F}_0,\d_{U^1}\mc{F}_0,\d_{U^2} \mc{F}_0,\d_{Z} \mc{F}_0\right)\,,\\
        \Omega_{1}&=\left(0,0,0,0,2\mc{F}_{1}-U^i\d_{U^i}\mc{F}_{1},\d_{U^1}\mc{F}_{1},\d_{U^2} \mc{F}_{1},\d_{Z} \mc{F}_{1}\right)\,,\\
        \tilde{\Omega}&=\Omega_0+\Omega_{1}\,.
    \end{aligned}
\end{equation}
$\tilde{\Omega}$ is the holomorphic 3-form corresponding to the newly defined $\tilde{\mc{F}}$, where the contributions from D3-brane wrapping modes have been removed. The new superpotential and K\"ahler potential read 
\begin{equation}
    \begin{aligned}
        W=&W_0+W_{\mc{C}}+W_{1}+\text{higher order terms}\,,\\
        W_0=&\int (F_3-\tau H_3)\wedge\Omega_0\,,\\
        W_{1}=&\int (F_3-\tau H_3)\wedge\Omega_{1}\,,
    \end{aligned}
\end{equation}
\begin{equation}
    \begin{aligned}
        \mc{K}=&\mc{K}_{0}+\mc{K}_{\mc{C}}+\text{higher order terms}\,,\\
        \mc{K}_{0}=&-\log \mc{V}_{\tilde{\Omega}}-\log \left(-i(\tau-\bar{\tau})\right)-2\log \mc{V}\,,\\
        \mc{V}_{\tilde{\Omega}}=&i \int\tilde{\Omega}\wedge\bar{\tilde{\Omega}}\,.
    \end{aligned}
\end{equation}
If we assume that all of the higher order corrections above are negligible, the above K\"ahler potential still has the no-scale property (\ref{no_scale_prop}), and the form (\ref{F_poten_orienti}) still applies. Hence to minimize the F-term potential, we should solve the  following equations
\begin{equation}
    D_{U^{\alpha}}W=D_Z W=D_{\tau}W=D_{\phi_a}W=0\,.
\end{equation}
Here $U^{\alpha},\alpha=1,2$ denote the complex structure moduli other than $Z$ which parametrizes the conifold singularity. We will also use the notation $U^i, i=1,2,3$ to denote all complex structure moduli with $U^3=Z$.

\subsection{Flux setup}

In order to do moduli stabilization, we need to choose a set of suitable fluxes. For the D3-brane wrapping modes to have a significant effect, we should require them to be light, i.e. $|Z|\ll1$. In order to achieve this, we choose a different flux configuration than in \cite{Alvarez-Garcia:2020pxd}, that fits in the DKMM construction \cite{Demirtas:2019sip}.

In general, we denote each component of the flux as 
\begin{equation}
    F_3=\left(\tilde{f}^0,\tilde{f}^i,f_0,f_i\right)^T,\; H_3=\left(\tilde{h}^0,\tilde{h}^i,h_0,h_i\right)^T\,.
\end{equation}
Similar to \cite{Alvarez-Garcia:2020pxd}, we can set
\begin{equation}
    F_3=\left(
    \begin{array}{c}
        0\\
         \tilde{f}^i\\
         b_i\tilde{f}^i \\
         \left(a_{i\alpha}\tilde{f}^i,f_3\right)^T\\
    \end{array}
    \right),\quad H_3=\left(
    \begin{array}{c}
        0\\
        0\\
         0  \\
        h_i\\
    \end{array}
    \right)\,.
\end{equation}
The flux should satisfy the following conditions:
\begin{itemize}
    \item $-\frac{1}{2}\vec{\tilde{f}}\cdot\vec{h}\leq Q_{D3}$.
    \item $N_{\alpha\beta}=K_{i\alpha\beta}\tilde{f}^i$ is invertible.
    \item $(N^{-1})^{\alpha\beta}h_{\alpha}h_{\beta}=0$.
    \item $p^{\alpha}=(N^{-1})^{\alpha\beta} h_{\beta}$ lies in the K\"ahler cone of the mirror CY.
    \item $a_{i\alpha}\tilde{f}^i$ and $b_i\tilde{f}^i$ are integer-valued.
\end{itemize}

In our CY3 $\mc{X}_6$ with $(h^{1,1},h^{2,1})=(99,3)$, these conditions imply the following relations between the components of $F_3$, $H_3$. By definition
\begin{equation}
    p^{\alpha}=\left(\frac{2\tilde{f}^2h_2-(2\tilde{f}^1+3\tilde{f}^2)h_1}{4(\tilde{f}^2)^2},\frac{h_1}{2\tilde{f}^2}\right)\,.
\end{equation}
In the DKMM construction \cite{Demirtas:2019sip}, to have a perturbatively flat vacuum there is a condition at the vacuum 
\be
\label{Ualphap}
U^{\alpha}=p^{\alpha}\tau\,.
\ee

As $(N^{-1})_{\alpha\beta}h^{\alpha}h^{\beta}=0\,$, from the expressions
\begin{equation}
    N=\left(
    \begin{array}{cc}
        0 & 2\tilde{f}^2 \\
        2\tilde{f}^2 & 2\tilde{f}^1+3\tilde{f}^2
    \end{array}
    \right),\quad  N^{-1}=\left(\begin{array}{cc}
        -\frac{2\tilde{f}^1+3\tilde{f}^2}{4(\tilde{f}^2)^2} & \frac{1}{2\tilde{f}^2} \\
       \frac{1}{2\tilde{f}^2}  & 0
    \end{array}\right)\,,
\end{equation}
we obtain an equation
\begin{equation}
    h_1\left[ 2 h_1 \tilde{f}^1+(3h_1-4h_2)\tilde{f}^2 \right]=0\,.
\end{equation}

To suppress the instanton prepotential (\ref{F_inst_eg}) we need $\op{Im}U^{\alpha}>0$, thus from (\ref{Ualphap}) $p^{\alpha}\neq 0$, and we require $h_1\neq 0$. Then
\begin{equation}
    2 h_1 \tilde{f}^1+(3h_1-4h_2)\tilde{f}^2 =0\,,
\end{equation}
and we can further simplify
\begin{equation}
    p^{\alpha}=\frac{1}{2\tilde{f}^2}(-h_2,h_1)\,.
\end{equation}
In order to make $\mc{F}_{\op{inst}}$ (\ref{F_inst_eg})  exponentially small, as $\op{Im}\tau=g_s^{-1}\gg 1$, from (\ref{Ualphap}) we need to assign $p^{\alpha}>0$. In the current case, as all triple intersection numbers $K_{ijk}$ are non-negative, it is straightforward to see that $p^{\alpha}$ always lies in the K\"ahler cone of the mirror CY. From the integrality conditions $a_{i\alpha}\tilde{f}^i, b_i\tilde{f}^i\in\mb{Z}$, it is required that $\tilde{f}^2=4 n_2$, $n_1,n_2\in \mb{Z}$.

Finally, our special requirement in this paper is that there is no flux through $\mc{C}$, i.e. $\tilde{f}^3=\tilde{h}^3=0$, to avoid changing the D3-brane wrapping modes. 

\subsection{Moduli stabilization}
\label{subsec:moduli_stabilize}

In the above flux setup, we can obtain the GVW superpotential
\begin{equation}
    \begin{aligned}
        W_0=&\left(F_3-\tau H_3\right)\cdot\Sigma\cdot \Omega_0\\
        =&\left(f_{\Lambda}U^{\Lambda}-\tilde{f}^{\Lambda}(\mc{F}_0)_{\Lambda}\right)-\tau \left(h_{\Lambda}U^{\Lambda}-\tilde{h}^{\Lambda}(\mc{F}_0)_{\Lambda}\right)\\
        =&(f_i-\tau h_i)U^i+f_0-\tau h_0 +\frac{1}{2}K_{ijk}(\tilde{f}^i-\tau\tilde{h^i})U^j U^k-a_{ij}(\tilde{f}^i-\tau\tilde{h}^i)U^j-(\tilde{f}^i-\tau \tilde{h}^i)b_i\\
        =&\frac{1}{2}K_{ijk}\tilde{f}^i U^j U^k-\tau h_iU^i + (f_3-a_{3i}\tilde{f}^i) Z\\
        =&\frac{1}{2}K_{i\alpha\beta}\tilde{f}^i U^{\alpha} U^{\beta}-\tau h_{\alpha}U^{\alpha} + (f_3-a_{3i}\tilde{f}^i+K_{i\alpha3}\tilde{f}^i U^{\alpha}-\tau h_3) Z+\frac{1}{2}K_{i33}\tilde{f}^i Z^2\,.
    \end{aligned}
\end{equation}
As we expect $Z$ is exponentially small with the same order of $W_1$, we can recombine each term in $W=W_U+W_Z+W_1$. The leading term is
\begin{equation}
    W_U=\frac{1}{2}K_{i\alpha\beta}\tilde{f}^i U^{\alpha} U^{\beta}-\tau h_{\alpha}U^{\alpha}\,,
\end{equation}
while the subleading terms are
\begin{equation}
    W_{Z}= Z\left[(f_3-a_{3i}\tilde{f}^i+K_{i\alpha3}\tilde{f}^i U^{\alpha}-\tau h_3)+\frac{1}{2}K_{i 33}\tilde{f}^{i} Z +\sqrt{2\tau}\phi_{1}\phi_{2}\right]\,,
\end{equation}
\begin{equation}
    \begin{aligned}
        W_{1}=&\int (F_3-\tau H_3)\wedge\Omega_{1}\\
        =&-\frac{1}{2\pi^2}\tilde{f}^1e^{2\pi i U^1}-\frac{63}{\pi^2}\tilde{f}^2 e^{2\pi i U^2}+ \text{ higher order}\,.
    \end{aligned}
\end{equation}
Then similar to \cite{Demirtas:2019sip}, we can approximate the value of $U^{\alpha}_0$ with $W_U$
\begin{equation}
    D_{U_{\alpha}}W\approx D_{U_{\alpha}}W_U= \d_{U_{\alpha}} W_U=N_{\alpha\beta}U^{\beta}-\tau h_{\alpha}=0\,,
    \label{DUalphaW}
\end{equation}
\begin{equation}
    U^{\alpha}=\tau (N^{-1})^{\alpha\beta}h_{\beta}= \tau p^{\alpha}\,.
\end{equation}
In the calculation we have used the result $W_U(U^{\alpha}_0)=0$, which results from the condition on flux $(N^{-1})^{\alpha\beta}h_{\alpha}h_{\beta}=0$. Hence the subleading terms should be of the same order $W\sim W_Z\sim W_1\sim\mc{O}(Z)$. As a consequence, when we calculate $D_{i} W$, we can approximate $\mc{V}_{\tilde{\Omega}}$ with $\mc{V}_{\Omega_0}$ in the term $W\d_{U^i}\mc{K}=-W\d_{U^i}\log\mc{V}_{\tilde{\Omega}}$ as the difference $\mc{V}_{\tilde{\Omega}}-\mc{V}_{\Omega_0}=\mc{O}(Z)$ is negligible.

When we take into account of the subleading terms, the above $U^{\alpha}$ receive a correction $U^{\alpha}=U^{\alpha}_0+\delta U^{\alpha}$. Here we will use the assumption $\delta U^{\alpha}\sim\mc{O}(Z)$, which can be crosschecked later.

For $D_{\tau}W$, it can be simplified using $h_\alpha U^\alpha=0$,
\begin{equation}
\label{DtauW}
    D_{\tau}W=-h_{\alpha}\delta U^{\alpha}-h_3 Z+\frac{1}{\sqrt{2\tau}}Z\phi_1\phi_2-\frac{1}{\tau-\bar{\tau}}W=0\,,
\end{equation}
\begin{equation}
    W=2i\op{Im}\tau(-h_{\alpha}\delta U^{\alpha}- h_3 Z+\frac{1}{\sqrt{2\tau}}Z\phi_1\phi_2)\,.
\end{equation}
After adding higher order terms $\delta U^{\alpha}$ in $D_{U^\alpha}W$, we have
\begin{equation}
    \begin{aligned}
    \label{DUalpha}
        D_{U^{\alpha}}W=&N_{\alpha\beta}U^{\beta}-\tau h_{\alpha}+K_{i\alpha3}\tilde{f}^iZ+\d_{U^{\alpha}}W_1-(\d_{U^{\alpha}}\log \mc{V}_{\Omega_0})W\,.
    \end{aligned}
\end{equation}
As we assume that $\delta U^{\alpha}= \mc{O}(Z)$, $\d_{U^{\alpha}}\log \mc{V}_{\Omega_0}$ can be approximated 
 with $U^{\alpha}\approx U^{\alpha}_0$. Then we can simplify the rest of the equations, $D_{\phi_{a}}W=0$, as
\begin{equation}
    \begin{aligned}
    \label{D12W}
        D_{\phi_1}W=\sqrt{2\tau}Z\phi_{2}+\kappa_4^2\phi_1^{\dagger}W&=0\,,\\
        D_{\phi_2}W=\sqrt{2\tau}Z\phi_{1}+\kappa_4^2\phi_2^{\dagger}W&=0\,.\\
    \end{aligned}
\end{equation}
In the following subsections, we will consider the following types of solutions to  $D_{\phi_a}W=0$: 
\begin{itemize}
    \item $Z\neq0,\phi_a\neq 0$, which corresponds to a vacuum in a new non-geometric phase.
    \item $Z=0,\phi_a\neq 0$, which shall be interpreted as the geometric phase of the resolved conifold $\tilde{\mc{X}}_6$.
    \item $Z\neq 0,\phi_a=0$, which corresponds to the geometric phase of the deformed conifold $\mc{X}_6$ with  $(h^{1,1},h^{2,1})=(99,3)$.
\end{itemize}

\subsubsection{Case $Z\neq 0, \phi_a\neq 0$}
From (\ref{D12W}), we obtain the relation
\begin{equation}
    |\phi_1|=|\phi_2|\,.
\end{equation}
We set
\begin{equation}
    \phi_1\phi_2=\phi^2e^{i\varphi},\; \phi\in \mb{R}_{\geq 0}\,,
\end{equation}
and then obtain the solved value for $W$
\begin{equation}
\label{Case1-Wsol}
    W=-\frac{1}{\kappa_4^2}e^{i\varphi}\sqrt{2\tau}Z\,.
\end{equation}
In our setup, we can only fix the absolute value $\phi$. Because we can always redefine $\phi_1\rightarrow\phi_1e^{i\theta}$, $\phi_2\rightarrow\phi_2e^{-i\theta}$ and obtain the same superpotential (\ref{superpoten_wrap_eg}) and K\"ahler potential (\ref{kahler_wrap_eg}), we are allowed to make the further assumption $\phi_1=\phi_2$, in order to obtain a reasonable physical approximation for the flux vacua and the quantum tunneling rate in the later sections.

We now expand $D_Z W$
\begin{equation}
    D_Z W=(f_3-a_{3i}\tilde{f}^i+K_{i\alpha3}\tilde{f}^i U^{\alpha}-\tau h_3+K_{i 33}\tilde{f}^{i} Z+\sqrt{2\tau}\phi^2e^{i\varphi})-(\d_Z\log \mc{V}_{\Omega_0})W=0\,,
\end{equation}
\begin{equation}
\label{DZW}
    (f_3-a_{3i}\tilde{f}^i+K_{i\alpha3}\tilde{f}^i U^{\alpha}-\tau h_3+\frac{1}{2}K_{i 33}\tilde{f}^{i} Z+\sqrt{2\tau}\phi^2e^{i\varphi})=\left(-\frac{1}{\kappa_4^2}e^{i\varphi}\sqrt{2\tau}(\d_Z\log \mc{V}_{\Omega_0})-\frac{1}{2}K_{i 33}\tilde{f}^{i} \right)Z\,.
\end{equation}
Combining with $D_{\tau} W$ (\ref{DtauW}), we get
\begin{equation}
    W=\frac{1}{2}N_{\alpha\beta}\delta U^{\alpha}\delta U^{\beta}+(f_3-a_{3i}\tilde{f}^i+K_{i\alpha3}\tilde{f}^i U^{\alpha}-\tau h_3+\frac{1}{2}K_{i 33}\tilde{f}^{i} Z+\sqrt{2\tau}\phi^2e^{i\varphi})Z+W_1=-\frac{1}{\kappa_4^2}e^{i\varphi}\sqrt{2\tau}Z\,,
\end{equation}
\begin{equation}
\label{SolveZ}
        \frac{1}{2}N_{\alpha\beta}\delta U^{\alpha}\delta U^{\beta}+\left(-\frac{1}{\kappa_4^2}e^{i\varphi}\sqrt{2\tau}(\d_Z\log \mc{V}_{\Omega_0})-\frac{1}{2}K_{i 33}\tilde{f}^{i} \right)Z^2+\frac{1}{\kappa_4^2}e^{i\varphi}\sqrt{2\tau}Z +W_1=0\,.   
\end{equation}
As $W_1$ only involves the exponential terms, and we have fixed $\op{Im}\tau$. If $2\pi\op{Im}U^{\alpha}\gg 1$, $|W_1|\ll 1$.
Ignore $\mc{O}(Z^2)$ terms, there is an exponentially small solution of $Z\ll 1$ to (\ref{SolveZ}):
\begin{equation}
    Z=  W_1 \frac{-\kappa_4^2}{e^{i\varphi}\sqrt{2\tau}}+\mc{O}(Z^2) \approx \frac{\kappa_4^2}{e^{i\varphi}\sqrt{2\tau}} \left(\frac{1}{2\pi^2}\tilde{f}^1e^{2\pi i U^1}+\frac{63}{\pi^2}\tilde{f}^2 e^{2\pi i U^2}\right)+\mc{O}(Z^2)\,.
\end{equation}
Then plug in (\ref{DZW}), we solve
\begin{equation}
    \phi^2=\left[\left(-\frac{1}{\kappa_4^2}e^{i\varphi}\sqrt{2\tau}(\d_Z\log \mc{V}_{\Omega_0})-K_{i 33}\tilde{f}^{i} \right)Z - (f_3-a_{3i}\tilde{f}^i+K_{i\alpha3}\tilde{f}^i U^{\alpha}-\tau h_3)\right]\frac{1}{e^{i\varphi}\sqrt{2\tau}}\in \mb{R}_{\geq 0}\,.
\end{equation}
When $f_3-a_{3i}\tilde{f}^i+K_{i\alpha3}\tilde{f}^i U^{\alpha}_0-\tau h_3\neq 0$, the $\mc{O}(Z)$ terms can be ignored, thus
\begin{equation}
    \phi^2=- (f_3-a_{3i}\tilde{f}^i+K_{i\alpha3}\tilde{f}^i U_0^{\alpha}-\tau h_3)\frac{1}{e^{i\varphi}\sqrt{2\tau}}\in \mb{R}_{\geq 0}\,.
\end{equation}
From the conditions on the flux components, we can find $K_{i\alpha 3}\tilde{f}^i p^\alpha=\frac{1}{2}h_1+h_2$, and simplify
\begin{equation}
    \sqrt{2\tau}\phi^2e^{i\varphi}=(h_3-\frac{1}{2}h_1-h_2)\tau-(f_3-a_{3i}\tilde{f}^i)\,.
\end{equation}
From $D_{U^{\alpha}}W=0$ (\ref{DUalpha}), we obtain
\begin{equation}
    \delta U^{\alpha}=(N^{-1})^{\alpha\beta}\left(-K_{i\beta3}\tilde{f}^i Z-\d_{U^{\alpha}}W_1+W\d_{U^{\beta}}\log \mc{V}_{\Omega_0}\right)\,.
\end{equation}
As in (\ref{Case1-Wsol}) we see $W\propto Z$, the above  expression means that $\delta U^{\alpha}\sim \mc{O}(Z)$, which is consistent with our assumption made before. Now combining $D_{\tau} W=0$ and $D_{\phi_{\zeta}}W=0$, with the equation
\begin{equation}
    i\op{Im}U^i \d_{U^i}\mc{V}_{\Omega_0}=\frac{3}{2}\mc{V}_{\Omega_0}+6\op{Im}\xi= i\op{Im}U^{\alpha} \d_{U^{\alpha}}\mc{V}_{\Omega_0}+\mc{O}(Z)\,,
\end{equation}
we arrive at
\begin{equation}
    W=-\frac{1}{\kappa_4^2}e^{i\varphi}\sqrt{2\tau}Z=2i\op{Im}\tau(-h_{\alpha}\delta U^{\alpha}- h_3 Z+\frac{1}{\sqrt{2\tau}}Z\phi^2e^{i\varphi})\,,
\end{equation}
\begin{equation}
    \begin{aligned}
        h_{\alpha}\delta U^{\alpha}=&-\frac{W}{2i\op{Im}\tau}- h_3 Z+\frac{1}{\sqrt{2\tau}}Z\phi^2e^{i\varphi}\\
        =&-\frac{W}{2i\op{Im}\tau}- h_3 Z+\frac{1}{2\tau}\left( (h_3-\frac{1}{2}h_1-h_2)\tau-(f_3-a_{3i}\tilde{f}^i) \right)Z\\
        =&-\frac{W}{2i\op{Im}\tau}- (\frac{1}{2}h_3+\frac{1}{4}h_1+\frac{1}{2}h_2)Z-\frac{f_3-a_{3i}\tilde{f}^i}{2\tau}Z\\
        =&p^{\alpha} \left(-K_{i\alpha 3}\tilde{f}^i Z-\d_{U^{\alpha}}W_1+W\d_{U^{\alpha}}\log \mc{V}_{\Omega_0}\right)\\
        =&-(\frac{1}{2}h_1+h_2)Z-p_{\alpha}\d_{U^{\alpha}}W_1 + W p^{\alpha}\d_{U^{\alpha}}\log\mc{V}_{\Omega_0}\,,
    \end{aligned}
\end{equation}
\begin{equation}
    \begin{aligned}
        &Z\left[ -i\op{Im}\tau\left( \frac{2h_3-h_1-2h_2}{4} + \frac{f_3-a_{3i}\tilde{f}^i}{2\tau} \right)   \right] \\
        =&\left( \frac{i\op{Im}U^{\alpha}\d_{U^{\alpha}}\mc{V}_{\Omega_0}}{\mc{V}_{\Omega_0}}+\frac{1}{2}\right)W+ip_{\alpha}\d_{U^{\alpha}}W_1\\
        =&\left(2+\frac{6\op{Im}\xi}{\mc{V}_{\Omega_0}}+\mc{O}(Z)\right)W+\mc{O}\left( \frac{W_1}{\op{Im\tau}} \right)\\
        =& (2+\mc{O}(\mc{V}_{\Omega_0}^{-1})+\mc{O}((\op{Im}\tau)^{-1})+\mc{O}(Z))W\,,
    \end{aligned}
\end{equation}
\begin{equation}
    i\op{Im}\tau\left( \frac{2h_3-h_1-2h_2}{4} + \frac{f_3-a_{3i}\tilde{f}^i}{2\tau} \right)\approx \frac{2}{\kappa_4^2}e^{i\varphi}\sqrt{2\tau}\,.
\end{equation}

For simplicity, we choose the flux to satisfy $f_3-a_{3i}\tilde{f}^i=0$, i.e. $f_3=-18$ in our example. Then we can set $e^{i\varphi}=e^{i\pi/4}$, i.e. $\tau=i\op{Im}\tau$, and obtain
\begin{equation}
    \op{Im}\tau \frac{2h_3-h_1-2h_2}{4}=\frac{2}{\kappa_4^2}\sqrt{2\op{Im}\tau}\in \mb{R}_+\,,
\end{equation}
Here we require that $2h_3-h_1-2h_2>0$. And finally we solve the imaginary part of the axiodilaton $\op{Im}\tau=g_s^{-1}$
\begin{equation}
    \op{Im}\tau=\left( \frac{8\sqrt{2}}{\kappa_4^2(2h_3-h_1-2h_2)} \right)^2\,.
\end{equation}

With this result, we should choose to satisfy $\frac{1}{2}h_1-h_3+h_2\neq 0$, to further solve the value for $\phi^2$
\begin{equation}
    \phi^2=- (f_3-a_{3i}\tilde{f}^i+K_{i\alpha3}\tilde{f}^i U_0^{\alpha}-\tau h_3)\frac{1}{e^{i\varphi}\sqrt{2\tau}}=(h_3-\frac{1}{2}h_1-h_2)\sqrt{\frac{\op{Im}\tau}{2}}\in \mb{R}_+\,.
\end{equation}
With the values for $\phi^2$, we find that (\ref{kahler_wrap_eg}) $\mc{K}_{\mc{C}}=\kappa_4^2(\phi_1\phi_1^{\dagger}+\phi_2\phi_2^{\dagger})\approx 8$\footnote{This result means that the absolute value of $\phi$ is considerably large, and there would be potentially large higher order corrections from the wrapping modes, to (\ref{kahler_wrap_eg},\ref{superpoten_wrap_eg}). However we do not carefully consider these unknown corrections in this work.}.


As a concrete example, we can set the following parameters
\be
\label{flux-param}
\tilde{f}^{i}=(-42,12,0)\ ,\ h_i=(1,-1,100)\ ,\ f_3=-18\ ,\ \mc{V}=10\ ,\ p^{\alpha}=(\frac{1}{24},\frac{1}{24})\,.
\ee
\begin{eqnarray}
    -\frac{1}{2}\vec{\tilde{f}}\cdot\vec{h}=27\leq Q_{D3}=52\,.
\end{eqnarray}
We solve the stabilized moduli fields and superpotential 
\begin{equation}
    \tau=5\times 10^1 i\,,\; U^i=(2 i, 2 i,-4\times10^{-7} i)\,,\; \phi= 2\times10^1\,,\; W=5\times 10^{-4}\,.
\end{equation}
Here we choose relatively small $\tilde{f}$, $h_1$, $h_2$, $\mc{V}$ and relatively large $h_3$ to obtain small $p^{\alpha}$ and $\op{Im}\tau$, which leads to relatively large $W$. From the result in Section \ref{sec:quantum-t}, we can see that small $\mc{V}$ and large $W$ will lead to a large tunneling rate.
Because we choose a not so large $\mc{V}=10$, it would be expected that the higher order corrections in the large volume scenario are not too small, hence we only keep one significant figure in our results.

\subsubsection{Case $Z=0,\phi_a\neq 0$}

Now we discuss the case corresponding to the resolved CY3 $\tilde{\mc{X}}_6$.
As we assume $\delta U^{\alpha}\propto Z$, in this case $\delta U^{\alpha}=0$. From $D_{\tau} W=0$ (\ref{DtauW}), 
\begin{equation}
    W=W_1\approx -\frac{1}{2\pi^2}\tilde{f}^1e^{2\pi i U^1}-\frac{63}{\pi^2}\tilde{f}^2 e^{2\pi i U^2}+ \frac{4626}{\pi^2}\tilde{f}^2e^{4\pi i U^2}=0\,.
\end{equation}
There is one unique solution 
\begin{equation}
    U^1=U^2=0.7 i\,.
\end{equation}
From $D_{U^{\alpha}}W$,
\begin{equation}
    D_{U^{\alpha}}W=\d_{U^{\alpha}}W=\d_{U^{\alpha}}W_U=0\,.
\end{equation}
Hence the exact solution is $U^{\alpha}=\tau p^{\alpha}$, which verifies the above assumption $\delta U^{\alpha}=0$. From $D_Z W$ (\ref{DZW}),
\begin{equation}
    D_Z W=f_3-a_{3i}\tilde{f}^i+K_{i\alpha3}\tilde{f}^i U^{\alpha}-\tau h_3+\sqrt{2\tau}\phi^2e^{i\varphi}=0\,.
\end{equation}
\begin{equation}
    \phi^2e^{i\varphi}=-\frac{1}{\sqrt{2\tau}}\left( f_3-a_{3i}\tilde{f}^i+K_{i\alpha3}\tilde{f}^i U^{\alpha}-\tau h_3 \right)\,.
\end{equation}
There is no other constraint on the $\phi_{a}$. To calculate the tunneling rate, similar to before we need to assume some relation, e.g. $\phi_1=\phi_2=\phi e^{i\varphi/2}$, then
\begin{equation}
    \phi e^{i\varphi/2}=\left[-\frac{1}{\sqrt{2\tau}}\left( f_3-a_{3i}\tilde{f}^i+K_{i\alpha3}\tilde{f}^i U^{\alpha}-\tau h_3 \right)\right]^{\frac{1}{2}}\,.
\end{equation}
When $f_3-a_{3i}\tilde{f}^i+K_{i\alpha3}\tilde{f}^i U^{\alpha}-\tau h_3=0$, we can reach the special solution $\phi=Z=0$. 

In order to realize quantum tunneling between vacua in the same flux setup, we choose the same parameters as the $\phi_a\neq 0, Z\neq 0$ case, 
\be
\tilde{f}^{i}=(-42,12,0)\ ,\ h_i=(1,-1,100)\ , \ f_3=-18\ ,\ \mc{V}=10\ ,\ p^{\alpha}=(\frac{1}{24},\frac{1}{24})\,,
\ee
We solve the values of the moduli fields and superpotential
\begin{equation}
    \tau=2\times 10 i\,,\; U^i=(0.7
    i, 0.7i,0)\,,\; \phi e^{i\varphi/2}= 2\times 10+7i\,,\; W=0\,.
\end{equation}

\subsubsection{Case $Z\neq 0,\phi_a= 0$}

This corresponds to the geometric phase of the deformed CY3 $\mc{X}_6$.

When $\phi_a=0$,  $W_{\mc{C}}=\d_{\tilde{M}}W_{\mc{C}}=0$ holds in the whole calculation. Thus we can obtain 
\begin{equation}
    D_{\tilde{M}}W=D_{\tilde{M}}(W_0+W_1)=0\,.
\end{equation}
In addition, in our setup, $\tilde{f}^0=\tilde{h}^0=0$, such that we have 
\begin{equation}
    W_0+W_1=\int (F_3-\tau H_3)\wedge\tilde{\Omega}=\int (F_3-\tau H_3)\wedge\Omega=W_{\op{flux}}\,.
\end{equation}
As a consequence, the whole analysis will be the same as in \cite{Demirtas:2019sip}. However, the previously chosen flux parameters (\ref{flux-param}) are inadequate to achieve moduli stabilization, due to the lack of flux through $\mc{C}$. To do moduli stabilization in this case we need to assign a different set of flux parameters, which would not be discussed in this paper as we are only interested in the different string vacua with the same flux background.

\section{Quantum tunneling rate}
\label{sec:quantum-t}

As shown in Section \ref{subsec:moduli_stabilize}, with a suitable choice of flux, we can obtain two vacua $\phi_a\neq 0, Z\neq 0$ and $\phi_a\neq 0, Z=0$ with the same set of flux parameters. According to the discussions in Section \ref{subsec:conifold_4dn=1}, these two vacua are in the non-geometric phase and the $\tilde{\mc{X}}_6$ phase respectively. In this section we discuss the quantum tunneling rate between such two phases, following the setups in  \cite{Coleman:1977py,Callan:1977pt,Devoto:2022qen,Coleman:1980aw,Parke:1982pm}.

Note that in this paper we do not provide an example of tunneling between $\mc{X}_6$ and $\tilde{\mc{X}}_6$, which is clearly a transition between different topology phases. In fact, as we mentioned above, it is hard to obtain two different vacua satisfying $\phi_a= 0, Z\neq 0$ and $\phi_a\neq 0, Z=0$ with a single set of flux parameters. Nonetheless, if we allow to shift between different choices of flux, as discussed in \cite{deAlwis:2013gka,Ahlqvist:2010ki,Danielsson:2006xw,Blanco-Pillado:2009lan}, we can realize such tunneling by two steps: the first step is to tunnel from one choice of flux vacua $(F_3,H_3)_1$ with $\phi_a= 0$, $Z\neq 0$ to another choice of flux vacua $(F_3,H_3)_2$ with $\phi_a\neq 0$, $Z=0$, and then we tunnel from $(F_3,H_3)_2$ with $\phi_a\neq 0$, $Z=0$ back to $(F_3,H_3)_1$ $\phi_a\neq 0$, $Z=0$ by changing the flux. We will not present the details of this more complicated quantum tunneling, and focus on the two vacua obtained in the last section.


\subsection{Vacuum bubble and thin wall approximation}

\subsubsection{Without gravity}

Let us first briefly review the tunneling without gravity \cite{Coleman:1977py}. Consider the 4d field theory with a single scalar $\phi$
\begin{equation}
    S=\int \dd^4 x-\frac{1}{2}\eta^{\hat{u}\hat{v}}\d_{\hat{u}}\phi\d_{\hat{v}}\phi-V(\phi)\,,\;\hat{u},\hat{v}=0,\dots 3\,.
\end{equation}
Assuming there are two vacua $\phi_t,\phi_f$ of $V(\phi)$, with $V(\phi_t)<V(\phi_f)=0$. We call $\phi_t$ the true vacuum and $\phi_f$ the false vacuum. When the whole universe is in the false vacuum $\phi_f$, it has a probability to decay to the true vacuum $\phi_t$ through some instanton transition $\phi_{\op{inst}}(x^{\hat{u}})$. Schematically, the decay rate is given by
\begin{equation}
\label{AB-decay}
    \Gamma=\ms{A}\exp\{-\ms{B}\}\,.
\end{equation}
$\ms{B}$ is the Euclidean action $S_E(\phi_{\op{inst}})$ of the instanton and the prefactor $\ms{A}$ can be determined by the semi-classical approximation $S_E(\phi)\approx S_E(\phi_{\op{inst}})+\frac{1}{2}\frac{\d^2 S_E}{\d c_{\mc{A}}\d c_{\mc{B}}}c_{\mc{A}}c_{\mc{B}}$ with $c_{\mc{A}}$ the moduli of the instanton.

When the temperature is small enough, i.e. $T\ll l_s^{-1}$, we can use the zero temperature approximation in our analysis. Then $\phi_{\op{inst}}$ is the solution of $S_E$ in $\mb{R}^4$ with $O(4)$ symmetry. With $O(4)$ symmetry, we simplify $\phi_{\op{inst}}(x^{\hat{u}})=\phi_{\op{inst}}(\rho)$, $x^{\hat{u}}\in \mb{R}^4$, $\rho\in \mb{R}_{\geq 0}$. The equation of motion for $\phi_{\op{inst}}$ in the radial direction is 
\begin{equation}
    \frac{\dd^2 \phi_{\op{inst}}}{\dd \rho^2}+\frac{3}{\rho}\frac{\dd \phi_{\op{inst}}}{\dd \rho}=V'(\phi_{\op{inst}})\,.
\end{equation}
As argued in \cite{Coleman:1977py}, such an equation always has a bounce solution $\phi_b(\rho)$ with finite $S_E(\phi_b)$
\begin{equation}
    \lim_{\rho\rightarrow\infty}\phi_b(\rho)=\phi_f\,,\;\left.\frac{\dd\phi_b(\rho)}{\dd \rho}\right|_{\rho=0}=0\,.
\end{equation}
In configuration, $\phi_b(\rho)$ is close to $\phi_f$ almost everywhere other than the region near $\rho=0$. It is called the vacuum bubble with the true vacuum inside the bubble and the false vacuum outside the bubble, similar to the boiling process of water.

When the potential difference $\delta V=V_f-V_t$ is small enough, we expect that the bubble wall is really thin, i.e. 
\begin{equation}
    \phi_b=\left\{
    \begin{aligned}
        &\phi_t\,,\;\rho< \rho_b\\
        &\phi_f\,,\;\rho>\rho_b+\Delta\rho
    \end{aligned}
    \right. \quad |\Delta\rho|\ll |\rho_b|\,.
\end{equation}
Then we can use the thin wall approximation, i.e. we can divide the whole action into three parts
\begin{eqnarray}
    \ms{B}=\ms{B}_{\op{in}}+\ms{B}_{\op{wall}}+\ms{B}_{\op{out}}\,.
\end{eqnarray}
Without gravity, these three parts can be approximated as \cite{Coleman:1977py}
\begin{equation}
    \begin{aligned}
    \ms{B}_{\text{n}}&=\ms{B}_{\text{int,n}}+\ms{B}_{\text{wall,n}}+\ms{B}_{\text{out,n}}\,,\\
        \ms{B}_{\text{int,n}}&=-\frac{1}{2}\pi^2\rho^4\epsilon\,,\\
        \ms{B}_{\text{wall,n}}&=2\pi^2\rho^3 \sigma\,,\\
        \ms{B}_{\text{out,n}}&=0\,.
    \end{aligned}
    \label{B_without_gravity}
\end{equation}
The difference between the vacuum energies is 
\begin{eqnarray}
    \epsilon=V_f-V_t>0\,.
\end{eqnarray}
In the case of a single real scalar, the bubble tension $\sigma$ is
\begin{equation}
    \sigma=\int_{\phi_t}^{\phi_f}\dd \phi\sqrt{2(V(\phi)-V_t)}\,.
\end{equation}
To maximize $\ms{B}_n$, we can find that 
\begin{eqnarray}
    \rho_n=\frac{3\sigma}{\epsilon}\,,
\end{eqnarray}
\begin{equation}
\label{O4-instanton}
\ms{B}_n \sim \ms{B}_0 \frac{\sigma^4}{\left(V_{f}-V_{t}\right)^3}\,.
\end{equation}
The prefactor can be computed as
\begin{equation}
    \ms{B}_0=\frac{27\pi^2}{2}\,.
\end{equation}
The thin-wall approximation is valid when 
\begin{equation}
    \sigma^4\gg \left(V_{f}-V_{t}\right)^3\,.
\end{equation}

\subsubsection{With gravity}

When gravity is strong, we need to modify the $4d$ action into 
\begin{equation}
    S=\int \dd^4 x \sqrt{-g}\left[\frac{1}{2\kappa_4^2}R-\frac{1}{2}g^{\hat{u}\hat{v}}\d_{\hat{u}}\phi\d_{\hat{v}}\phi-V(\phi)\right]\,,\;\hat{u},\hat{v}=0,\dots 3\,.
\end{equation}
Here the metric $g$ is the graviton field. Once again, we can compute the tunneling between the two vacua by the action of the instanton. We can still use the thin wall approximation with 
\begin{equation}
    \begin{aligned}
    \ms{B}_{\text{g}}&=\ms{B}_{\text{int,g}}+\ms{B}_{\text{wall,g}}+\ms{B}_{\text{out,g}}\,,\\
        \ms{B}_{\text{int,g}}&=\frac{12\pi^2}{\kappa_4^4}\left[\frac{\left(1-\frac{1}{3}\kappa_4^2\rho^2 V_t\right)^{\frac{3}{2}}-1}{V_t}-\frac{\left(1-\frac{1}{3}\kappa_4^2\rho^2 V_f\right)^{\frac{3}{2}}-1}{V_f}\right]\,,\\
        \ms{B}_{\text{wall,g}}&=2\pi^2\rho^3 \sigma\,,\\
        \ms{B}_{\text{out,g}}&=0\,.
    \end{aligned}
    \label{B_with_gravity}
\end{equation}
When $\kappa_4\rightarrow 0$ while $\sigma$ and $\epsilon$ keep finite, this equation is reduced to the case without gravity (\ref{B_without_gravity}). 

Once again, we need to find the maximum of this action. 
As we require that the Euclidean action $B$ is real and the radius of the bubble $\rho\geq 0$, sometimes we can not find the maximum by solving the equation $\frac{\dd \ms{B}_g}{\dd \rho}=0$. In the case where $V_f\leq 0$, we can take $\rho\rightarrow\infty$. In this limit, 
\begin{equation}
    \ms{B}_g\rightarrow \left[2\pi^2\sigma  -\frac{4\pi^2}{\sqrt{3}\kappa_4}\left( \sqrt{-V_t}-\sqrt{-V_f}\right)\right]\rho^3\,.
\end{equation}
So we can find that when
\begin{eqnarray}
    \frac{\sqrt{3}}{2}\sigma>\frac{\sqrt{-V_t}-\sqrt{-V_f}}{\kappa_4}\,,
\end{eqnarray}
the maximum locates at $\rho\rightarrow \infty$ and $\ms{B}_g\rightarrow\infty$, $\Gamma\rightarrow 0$. So in this case, when $V_f\leq 0$ the tunneling is prevented.
When $V_f>0$, as $\ms{B}_g$ is real, there is a bound
\begin{equation}
    \rho\leq \sqrt{\frac{3}{\kappa^2_4V_f}}\,.
\end{equation}
Then the maximum locates at this bound and 
\begin{equation}
    \ms{B}_g\leq 2\pi^2\sigma\left(\frac{3}{\kappa_4^2 V_f}\right)^{\frac{3}{2}}+\frac{12\pi^2}{\kappa_4^4}\left[\frac{(1-\frac{V_t}{V_f})^{\frac{3}{2}}-1}{V_t}+\frac{1}{V_f}\right]\,.
    \label{Bg_bound_dS}
\end{equation}
Such a bound can not be the action of the true instanton as this is not a stable configuration with zero derivatives. But we can use this expression to obtain an approximate lower bound of the tunneling rate.

When 
\begin{eqnarray}
    \frac{\sqrt{3}}{2}\sigma<\frac{\sqrt{-V_t}-\sqrt{-V_f}}{\kappa_4}\,,
\end{eqnarray}
there exists a local maximum at finite $\rho>0$.
As shown in \cite{Coleman:1980aw,Parke:1982pm}, 
\begin{equation}
    \rho_g=\frac{\rho_n}{\sqrt{1+2\left(\frac{\rho_n}{2\lambda}\right)^2+\left(\frac{\rho_n}{2\Lambda}\right)^4}}\,,
\end{equation}
\begin{equation}
        \lambda^2=\frac{3}{\kappa_4^2(V_f+V_t)}\,,\;\;\Lambda^2=\frac{3}{\kappa_4^2(V_f-V_t)}\,.
\end{equation}
\begin{equation}
    \ms{B}_g=\ms{B}_n r\left(\frac{\rho_n^2}{4\lambda^2},\frac{\lambda^2}{\Lambda^2}\right)\,,\;\;r(x,y)=\frac{2\left[(1+3x+2x^2+x^2y^2+x^3y^2)-(1+2x+x^2y^2)^{\frac{3}{2}}\right]}{x^2(1-y^2)(1+2x+x^2y^2)^{\frac{3}{2}}}\,.
\end{equation}

\subsection{Quantum tunneling among different phases}

In our string compactification setup, it is difficult to find the exact instanton solution between the two vacua $\phi_a\neq 0,Z\neq 0$ and $\phi_a\neq 0,Z= 0$. We will instead use the thin wall approximation, which only requires the information of the scalar potential. We will denote the local minimum with $\phi_a\neq 0,Z\neq 0$ as the true vacuum $\phi_{\tilde{M},t}$ and the local minimum $\phi_a\neq 0,Z= 0$ as the false vacuum $\phi_{\tilde{M},f}$. Here $\tilde{M}$ ranges over all the complex scalars $U^i,\tau,\phi_a$.

The expression of the tunneling rate \cite{Coleman:1977py} is only valid for the case of a single real scalar, which is quite different from the string compactification scenario with multiple complex scalars. In the following computations, we only keep track of the orders of magnitude. 

Similar to \cite{Ahlqvist:2010ki}, we can choose a simple path $\phi_{\tilde{M}}(t)=\phi_{\tilde{M},f}+(\phi_{\tilde{M},t}-\phi_{\tilde{M},f})t$, $t\in [0,1]$, connecting the two vacua in Section \ref{subsec:moduli_stabilize}. We can use this path to transform the whole supergravity into an effective field theory with one real scalar and use this effective theory to approximate $\ms{B}$. Then the bubble tension is approximated as
\begin{equation}
    \sigma=\int_0^1\dd t \sqrt{\mc{K}_{\tilde{M}\tilde{N}}\dd\phi^{\tilde{M}}\dd\phi^{\tilde{N}}}\sqrt{2(V[\phi]-V_t)}\approx 10^2\,.
\end{equation}
In the CY compactification, there are multiple scalar fields. So in principle we need to compute the action $\ms{B}_{\op{total}}$ of the instanton in the whole supergravity, which should be less than the result from the single field approximation $\ms{B}_{\op{total}}\leq \ms{B}_{\op{single}}$. But in our work, we focus on the upper bound of $\ms{B}$. For simplicity, we will only use the single field method to approximate $\ms{B}\lesssim \ms{B}_{\op{single}}$. For the value of $\ms{B}_0$, we assume that it is still of a similar order to the single field approximation
\begin{equation}
    \ms{B}_0\sim \mc{O}(10^2)\,.
\end{equation}
When there are multiple fields, we can expect the order of $\ms{B}_0$ will not change significantly. So this approximation will not change our final approximation significantly.

\begin{figure}
    \centering
    \includegraphics[width=0.5\linewidth]{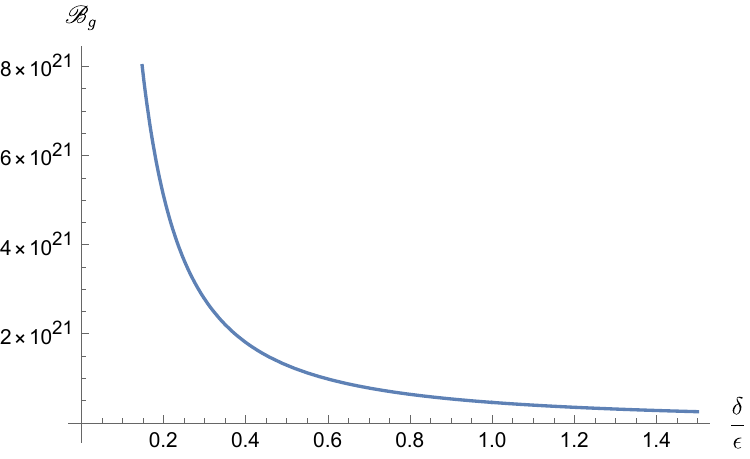}
    \caption{The upper bound of $\ms{B}_g$ with the uplift $\delta$ to $V_f$ and $V_t$. The horizontal axis is the ratio $\delta/\epsilon$, where $\epsilon=V_f-V_t$.}
    \label{fig:Bg_bound}
\end{figure}

As for the potential, from the moduli stabilization process, we know that at the tree level
\begin{equation}
    V_F(\phi_{\tilde{M},f})=V_F(\phi_{\tilde{M},t})=0\,.
\end{equation}
In order to have a difference between $V_f$ and $V_t$, we must take into account of the subleading correction of the potential. In our setup, in the false vacuum, $W(\phi_f)=0$, so the subleading correction $\delta V_f=0$. As for the true vacuum, we estimated that the subleading correction majorly comes from the term (\ref{deltaV-loop}) (note that $\kappa_4=\sqrt{1/4\pi\mc{V}}\sim 10^{-1}$):
\be
\ba
    \delta V_{\rm loop}&\sim  - \frac{|W_0|^2}{4 \pi \kappa^2_4} \mc{O}(\frac{1}{\mc{V}^{10/3}})
    \sim -\mc{O}(10^{-9})\,.
\ea
\ee
Hence obtain the difference in scalar potential
\begin{equation}
    \delta V=\delta V_f-\delta V_t\sim \mc{O}(10^{-9})\,,
\end{equation}
satisfying the thin wall approximation condition
\begin{equation}
    \sigma^4\sim \mc{O}(10^8)\gg \left(V_{f}-V_{t}\right)^3\sim \mc{O}(10^{-27})\,.
\end{equation}
After a crude estimation of $\ms{A}\sim\mc{O}(1)$ in (\ref{AB-decay}), we can obtain 
\be
    \ms{B}_n\sim \mc{O}(10^{33})\,.
\ee
If we take into account of the gravity effect, we can find that in our case
\begin{equation}
    \frac{\sqrt{3}}{2}\sigma\gg \frac{\sqrt{-V_t}-\sqrt{-V_f}}{\kappa_4}\,.
\end{equation}
So in our construction, the tunneling between the Minkowski vacuum and the AdS vacuum is prevented. However, if there exists some uplift to the vacuum energy $V_f$, e.g. a small uplift from anti-D3-branes \cite{Kachru:2003aw, Klebanov:2000hb} or other mechanisms~\cite{Cicoli:2015ylx}, tunneling is possible, while the bound (\ref{Bg_bound_dS}) is highly dependent on the uplift. We can expect there exists an uplift $\delta$ on both vacua 
\begin{eqnarray}
    V_f\rightarrow V_f+\delta=\delta\,,\quad V_t\rightarrow V_t+\delta=\delta-\epsilon\,.
\end{eqnarray}
Then we can obtain the bound
\begin{equation}
    \ms{B}_g=2\pi^2\sigma\left(\frac{3}{\kappa_4^2}\right)^{\frac{3}{2}}\delta^{-\frac{3}{2}}+\frac{12\pi^2}{\kappa_4^4}\left[ \frac{\epsilon^{\frac{3}{2}}-\delta^{\frac{3}{2}}}{(\delta-\epsilon)\delta^{\frac{3}{2}}}+\frac{1}{\delta} \right]\,.
\end{equation}
We plot the dependence of the upper bound of $\ms{B}_g$ on $\delta$ in Figure~\ref{fig:Bg_bound}. Obviously, when $\frac{\delta}{\epsilon}\sim \mc{O}(1)$, we can find that $\ms{B}_g\sim \mc{O}(10^{21})\ll \ms{B}_n\sim \mc{O}(10^{33})$. So we can find that the gravitational effect suppresses $\ms{B}$ and leads to a much larger tunneling rate.

However, in Planck units, the age of the universe is approximately
\begin{equation}
    \log T_{\op{universe}}\sim \mc{O}(10^2)\,,
\end{equation}
which is much smaller than $\ms{B}$. Hence the quantum tunneling rate between different phases is still too small to be observed in the history of the universe. But they may have significance in scenarios of eternal inflation~\cite{Hebecker:2020aqr}.

\section{Caveats and discussions}
\label{sec:discussions}

In this work, we analyzed the vacuum tunneling process from the conifold transitions in type IIB CY3 orientifold compactifications. On the complex deformation side $\mc{X}_6$, near the conifold singularity, there is a series of BPS hypermultiplets which become massless at the conifold singularity, from the D3-branes wrapping the vanishing 3-cycles. By including the explicit dependence of such a part in the 4d $\mc{N}=1$ supergravity, we can understand the phase transition among $\mc{X}_6$, $\tilde{\mc{X}}_6$ and a new non-geometric phase.

Through the semi-classical quantization of the D3-brane wrapping modes, we proposed the possible forms of the K\"ahler potential and the superpotential. Then we included these additional contributions from wrapping modes to the K\"ahler potential and the superpotential of the orientifold compactification, and also use the supersymmetry to constrain the interaction terms between the geometric moduli and the D3-brane wrapping modes. 

With this new supergravity action, we performed the moduli stabilization with a suitable choice of flux that does not pass through the vanishing 3-cycle $\mc{C}$. We obtained the local minima of the scalar potential $V_F$ with $\phi_a\neq 0, Z\neq 0$ and $\phi_a\neq 0, Z= 0$, i.e. the non-geometric phase and the $\tilde{\mc{X}}_6$ phase with the same set of flux parameters. These two vacua are treated as the true vacuum and the false vacuum respectively, in the local region of the landscape. Note that due to the reduction of supersymmetry from $\mc{N}=2$ to $\mc{N}=1$, as well as potential large values of $\phi_a$ comparing to the CY3 volume $\mc{V}$ (to try to increase the vacuum decay rate), the supergravity action is subject to other higher order corrections. Thus our analysis is crude and should be interpreted as giving a rough estimation of the vacuum decay rate.

Finally, with a simple model of the vacuum bubble and the thin wall approximation, we approximate the exponent $\ms{B}$ of the quantum tunneling rate between the true vacuum and the false vacuum. We find that with the common assumptions of large volume, weak coupling and near Minkowski 4d metric, we would obtain an extremely small tunneling rate. In other words, the tunneling process between different phases can only be effectively observed in the strong coupling region of string theory, which is beyond the scope of current understanding of string theory. 

Hence if we assume that the string landscape picture describes the real  universe, which is currently in the aforementioned large volume, weak coupling, and near Minkowski regime, the topology of the compact space can be considered as fixed.

However, in our work, we only focused on the simplest case where there is no flux through the vanishing 3-cycles. We also only discussed the transition from the complex deformation side $\mc{X}_6$ and left out the resolved side $\tilde{\mc{X}}_6$. In the following part, we will give a brief discussion about the difficulties we faced in these parts.

\subsection{Flux through the vanishing cycles}

In Appendix \ref{sec:BPS_3-cycle}, we review the derivation of the BPS condition of the 3-cycles. However, this analysis is only valid when there is no flux through the vanishing cycles. In fact, there is a rather confusing result in \cite{Marino:1999af} that when there is non-trivial flux through the 3-cycle, a D3-brane wrapping on the 3-cycle cannot be supersymmetric. While our discussion is highly dependent on the special Lagrangian condition and the supersymmetry, the cases of non-trivial flux will face additional complications.

In addition, in our discussion, we only take into account of the massive wrapping modes whose masses are proportional to the volumes of the 3-cycles. But when there is non-trivial flux through the vanishing 3-cycles, there will be additional KK modes lighter than the wrapping modes from the warped geometry \cite{Blumenhagen:2019qcg}. These KK modes should be included in the effective field theory as well in addition to the wrapping modes, which makes the theory much more complicated.

\subsection{Difficulty on the resolved side}

Comparing to the deformed CY3 $\mc{X}_6$, the physical analysis on the resolved CY3 $\tilde{\mc{X}}_6$ is much more involved. One reason is that on the resolved CY3 $\tilde{\mc{X}}_6$, there are vanishing $n_{\mc{C}}$ 2-cycles instead of $n_{\mc{C}}$ 3-cycles. Hence at the conifold singularity, from D3-branes wrapping the vanishing cycles, we obtain the $n_{\mc{C}}$ additional tensionless strings instead of the  $n_{\mc{C}}$ additional massless hypermultiplets. The quantization of such tensionless strings and the resulted effective action is a long-known problem. It is also curious to understand why the physical effects of $n_{\mc{C}}$ tensionless strings can be treated as equivalent to the $n_{\mc{C}}-n_{L}$ hypermultiplets and vector multiplets.

We can also see such a problem from the quantum correction perspective. In the complex deformation side, the worldsheet instanton correction in the mirror manifold generates the correction (\ref{F_inst_general}). Near the conifold singularity, there are only finite many terms become dominant as one $U\rightarrow 0$, so they can be generated from finite many degrees of freedom, i.e. the wrapping modes.

On the resolved side, the world-sheet instanton correction can generate the correction \cite{Dine:1986zy,Dine:1987bq}
\begin{equation}
\delta \mathcal{V}_{\mathrm{WSI}}=\frac{1}{2(2 \pi)^3} \sum_{\mathbf{q} \in \mathcal{M}(X)} \mathscr{N}_{\mathbf{q}}\left(\operatorname{Li}_3\left((-1)^{\gamma \cdot \mathbf{q}} e^{-2 \pi \mathbf{q} \cdot \mathbf{t}}\right)+2 \pi \mathbf{q} \cdot \mathbf{t} \operatorname{Li}_2\left((-1)^{\gamma \cdot \mathbf{q}} e^{-2 \pi \mathbf{q} \cdot \mathbf{t}}\right)\right)\,.
\end{equation}
From the discussion in \cite{Ooguri:1996me}, near the conifold singularity there are infinitely many terms becoming dominant as one $T\rightarrow 0$. These terms corresponding to infinitely many degrees of freedom, i.e. the SCFT sector from the tensionless strings. Such a phenomenon also implies that it is almost impossible to simply attach the resolved side and the complex deformation side together and discuss the tunneling through the attached theory.

In summary, on the deformation side, we can describe the phase transition perturbatively by introducing finite supermultiplets. But on the resolved side, non-perturbative effects are dominant and we can not simply use the Higgs mechanism on the classical level to realize the phase transition. In fact, this implies that there is a strong-weak duality between the resolved side and the deformation side. And we will only focus on the weak coupling description.


\subsection{Possibility of large tunneling rates}

In our construction, $\sigma\sim \mc{O}(10^2)$ and $\delta V$ is exponentially small. So even after the uplift, the tunneling rate is too small to observe such a tunneling process in the real universe. If we want to realize tunneling rates comparable with the age of the universe, i.e. $\mathscr{B}\sim 10^2$. This can be achieved by some larger $\delta V\sim \sigma\gtrsim 1$ or smaller $\sigma\ll 1$.

 In our construction, the subleading term $\delta V_{\op{loop}}$ is suppressed by the large volume $\mc{V}$,  the small superpotential $|W_0|$. If we want to achieve high tunneling rates, we need to work in a regime where $\mc{V}$, $|W|$ are large enough such that $\frac{\sigma^4}{\delta V_{\op{loop}}^3}\gtrsim 1$. This corresponds to a subset of string vacua with small volume or a large deviation from Minkowski spacetime. To make the supergravity description reliable, we should keep the large volume and the weak coupling condition. So we can only realize large the $\delta V$ through the large $|W_0|$. In our case, which is similar to the KKLT scenario\cite{Demirtas:2019sip,Blumenhagen:2019qcg,Alvarez-Garcia:2020pxd}, this is almost impossible because $\log |W_0|\sim -g_s^{-1}$. However, in the LVS scenario, we can realize vacua with large $|W_0|$ \cite{Balasubramanian:2005zx, Conlon:2005ki, Cicoli:2008va}. So, it might be possible to find some vacua with large tunneling rates.

 As for the small $\sigma$, this means that the distance between the two vacua is exponentially small.  In our construction, the distance between the two vacua is of finite value in general. So we need to find some new constructions to realize such a small distance case.

From the above discussion, we can find why we use the quantum corrected $\mc{V}$ instead of the classical volume $\mc{V}_0$ of $\mc{X}_6$ in the mass of the wrapping modes (\ref{3-cylce_volume_corr}). Without incorporating this correction, the leading contribution to the higher order potential $\delta V$ would be given by an additional (\ref{deltaV-leading}) rather than (\ref{deltaV-loop}). (\ref{deltaV-leading}) can be relatively large in the weak coupling limit, which implies that in the weak coupling limit, the tunneling rate between different phases in $\mc{M}_E$ becomes relatively large, contradicting physical intuition. In the LVS, when the string coupling enters the strongly coupled regime, the various quantum corrections discussed in Section (\ref{subsec:higherorder}) begin to compete with each other. Then   the logarithmic enhancement coming from the marginal operator in the D-brane/O-plane system  (\ref{deltaV-marginal}) may dominate the contribution, potentially leading to large tunneling rates.



\acknowledgments
We would like to thank Haipeng An, Bin Chen, Kentaro Hori, Zhiwei Li, Jianxin Lu, Yi Wang, Taizan Watari, Peihe Yang, Jingjun Zhang for helpful discussions. XG is supported by the National Natural Science Foundation of China under Grant No. 12375065. QJL and YNW are supported by the National Natural Science Foundation of China under Grant No. 12175004, No. 12422503 and by the Young Elite Scientists Sponsorship Program by CAST (2024QNRC001).

\appendix

\section{F-term potential in the Einstein frame}
\label{sec:F_potential}

In \cite{Conlon:2005ki}, in the modified Einstein frame with metric $G_e$, i.e. $G_s=e^{(\Phi-\Phi_0)/2}G_e=e^{\Phi/2}G_E$, $G_e=g_s^{\frac{1}{2}}G_E$, the potential is
\begin{equation}
    \int M_p^2V_F *1=\frac{g_s^4 M_{p,e}^4}{8\pi \mc{V}_e^2}\int \dd^4 x\sqrt{-G_e}e^{-\log\mc{V}_{\Omega}}\left(\mc{K}^{MN} D_MW D_{\bar{N}}\bar{W}-3|W|^2\right)\,,
\end{equation}

To transform from the modified Einstein frame to the Einstein frame, we use the following result
\begin{equation}
    \mc{V}_e=\int\dd^6 x\sqrt{-G_e}=g_s^{\frac{3}{2}}\int\dd^6 x\sqrt{-G_E}=g_s^{\frac{3}{2}}\mc{V}\,,
\end{equation}
\begin{equation}
    \int \dd^4 x\sqrt{-G_e}=g_s\int \dd^4 x\sqrt{-G_E}\,,
\end{equation}
\begin{equation}
    M_{p,e}^2=\frac{4\pi\mc{V}_e}{g_s^2 l_s^2}=\frac{4\pi\mc{V}}{g_s^{\frac{1}{2}} l_s^2}=g_s^{-\frac{1}{2}}M_p^2\,,
\end{equation}
\begin{equation}
    e^{\mc{K}}=e^{-\log \mc{V}_{\Omega}-2\log\mc{V}-\log(-i(\tau-\bar{\tau}))}=e^{-\log\mc{V}_{\Omega}}\frac{g_s}{2\mc{V}^2}\,.
\end{equation}
For simplicity, we keep the convention
\begin{equation}
    W_{\op{flux}}=\frac{1}{l_s^2}\int G_3\wedge\Omega\,.
\end{equation}
Then we can obtain
\begin{equation}
    V_F=\frac{1}{4\pi\kappa_4^2}e^{\mc{K}}\left(\mc{K}^{M\bar{N}} D_{M}W D_{\bar{N}}\bar{W}-3|W|^2\right)\,.
\end{equation}

\section{BPS 3-cycles and the volume}
\label{sec:BPS_3-cycle}

Here we review the process of computing the volume of the BPS 3-cycle in \cite{Becker:1995kb,Becker:2006dvp}. We argue that the BPS conditions for Euclidean M2-branes and D3-branes wrapping supersymmetric 3-cycles are the same, which are unique up to a convention as mentioned in \cite{Becker:1995kb}.

With the notations in \cite{Cicoli:2009zh}, in a Calabi-Yau threefold, there is a global spinor $\eta=e^{i\theta}\eta_{+}+e^{-i\theta}\eta_{-}$, where $\eta_{\pm}$ are components with different chiralities. Choosing the normalization of the spinor $\eta^{\dagger}_{\pm}\eta_{\pm}=1$ in this convention, we can construct
\begin{equation}
    \eta_{ \pm}^{\dagger} \gamma^{m n} \eta_{ \pm}= \pm i J^{m n}, \quad \eta_{-}^{\dagger} \gamma^{m n p} \eta_{+}=i \Omega^{m n p}, \quad \eta_{+}^{\dagger} \gamma^{m n p} \eta_{-}= - i \bar{\Omega}^{m n p}\,,
\end{equation}
\begin{equation}
    \gamma^{m_1 \ldots m_p}=\gamma^{\left[m_1\right.} \gamma^{m_2} \ldots \gamma^{\left.m_p\right]}\,,
\end{equation}
\begin{equation}
J \wedge J \wedge J=\frac{3 i}{4} \Omega \wedge \bar{\Omega}, \quad J \wedge \Omega=0\,.
\end{equation}
The holomorphic $(3,0)$-form $\Omega$ is defined up to a scaling factor, hence we can generally set
\begin{equation}
    \Omega = e^{\mc{A}+i\mc{B}}\eta_{-}^{\dagger} \gamma^{m n p} \eta_{+}, \quad \mc{A},\mc{B}\in\mb{R}\,.
    \label{holo_form_with_scale}
\end{equation}
In this new convention, we can find that
\begin{equation}
    J \wedge J \wedge J=\frac{3 i}{4} e^{-2\mc{A}}\Omega \wedge \bar{\Omega},\quad \mc{V}_{\Omega}=8 e^{2\mc{A}} \mc{V}_0\,,
\end{equation}
\begin{equation}
    \mc{V}_0=\frac{1}{6}\int J\wedge J\wedge J,\quad \mc{V}_{\Omega}=i\int\Omega\wedge\bar{\Omega}\,.
\end{equation}
\begin{equation}
    \mc{A}=\frac{1}{2}\left(\log \mc{V}_{\Omega}-\log 8\mc{V}_0\right)\,.
\end{equation}
This assignment of $\mc{A}$ is unique, instead of just a convention.

Then we can discuss the volume of the BPS 3-cycle. As shown in \cite{Becker:1995kb,Marino:1999af}, in type IIB, the BPS condition for the D3-brane is that there exists a global spinor $\eta$,
\begin{equation}
    \delta_{\eta}\Theta+\delta_{\kappa}\Theta=0\,.
\end{equation}
Here the two terms are the susy variation and the $\kappa$-symmetry variation
\begin{equation}
    \delta_{\eta}\Theta=\eta,\quad \delta_{\kappa}\Theta=2P_{+}\kappa(\theta)\,.
\end{equation}
When there is no flux through the D3-brane, the projection operators are
\begin{equation}
    P_{\pm}=\frac{1}{2}\left( 1\pm \frac{i}{4 !}\frac{\epsilon^{\mu\nu\rho\sigma}}{\sqrt{-h}}\d_{\mu}X^I\d_{\nu}X^J\d_{\rho}X^K\d_{\sigma}X^L\,\sigma_2\otimes\Gamma_{IJKL}\right),\quad (P_{\pm})^2=P_{\pm},\quad P_+P_-=0\,.
\end{equation}
Here $\sigma_2$ acts on the two supercharges in type IIB. As for the D3-brane wrapping a compact 3-cycle $\mc{C}$, i.e. the worldvolume of the D3-brane is $\mc{T\times C}$. Then we can simplify
\begin{equation}
    \begin{aligned}
        \frac{i}{4 !}\frac{\epsilon^{\mu\nu\rho\sigma}}{\sqrt{-h}}\d_{\mu}X^I\d_{\nu}X^J\d_{\rho}X^K\d_{\sigma}X^L\,\sigma_2\otimes\Gamma_{IJKL}=&\frac{i}{3 !}\frac{\epsilon^{0\hat{\mu}\hat{\nu}\hat{\rho} }}{\sqrt{-h}}\d_{0}X^0\d_{\hat{\mu}}X^m\d_{\hat{\nu}}X^n\d_{\hat{\rho}}X^p\,\sigma_2\otimes\Gamma_0\Gamma_{mnp}\\
        =&\frac{i}{3 !}\frac{\epsilon^{\hat{\mu}\hat{\nu}\hat{\rho} }}{\sqrt{-h}}\d_{\hat{\mu}}X^m\d_{\hat{\nu}}X^n\d_{\hat{\rho}}X^p\,\sigma_2\otimes\Gamma_0\Gamma_{mnp}\,.
    \end{aligned}
\end{equation}
As we only focus on one of the supercharges in type IIB, the Pauli matrix factor $\sigma_2$ can be ignored. We decompose the 10d spinor  into the spinors in $M_4$ and $\mc{X}_6$
\begin{equation}
\varepsilon=\lambda \otimes \eta_{+}+\lambda^* \otimes \eta_{-}\,,
\end{equation}
and only focus on the spinors $\eta$ in $\mc{X}_6$. On $\mc{X}_6$ we  reduce the 10d gamma matrix $\Gamma_m$ into the 6d gamma matrix $\gamma_m$ in $\mc{X}_6$ by removing the factor $\Gamma_0$. Consequently we obtain the same form of projectors as the cases of Euclidean M2-branes in \cite{Becker:1995kb}
\begin{equation}
    P_{\pm}=\frac{1}{2}\left( 1\pm \frac{i}{3 !}\frac{\epsilon^{\hat{\mu}\hat{\nu}\hat{\rho} }}{\sqrt{-h}}\d_{\hat{\mu}}X^m\d_{\hat{\nu}}X^n\d_{\hat{\rho}}X^p\,\gamma_{mnp}\right),\quad (P_{\pm})^2=P_{\pm},\quad P_+P_-=0\,.
\end{equation}
The supersymmetry condition on $\eta$ reads
\begin{equation}
    P_- \eta=0\,.
\end{equation}
Using the holomorphic coordinates $X^{\hat{m}}$, $X^{\bar{\hat{m}}}$, as $\gamma_{\hat{m}}\eta_+=0$, we get  $\gamma_{\hat{m}\hat{n}\hat{p}}\eta_+=0$,$\gamma_{\hat{m}\hat{n}\bar{\hat{p}}}\eta_+=0$. Then the non-zero terms are $\gamma_{\hat{m}\bar{\hat{n}}\bar{\hat{p}}}\eta_+$, $\gamma_{\bar{\hat{m}}\bar{\hat{n}}\bar{\hat{p}}}\eta_+$. As $\left\{\gamma_{\hat{m}},\gamma_{\bar{\hat{n}}}\right\}=G_{\hat{m}\bar{\hat{n}}}$ and $J_{\hat{m}\bar{\hat{n}}}=iG_{\hat{m}\bar{\hat{n}}}$, we find 
\begin{equation}
    \gamma_{\hat{m}\bar{\hat{n}}\bar{\hat{p}}}\eta_+=-2iJ_{\hat{m}[\bar{\hat{n}}}\gamma_{\bar{\hat{p}}]}\eta_+\,.
\end{equation}
Together with (\ref{holo_form_with_scale}), we obtain
\begin{equation}
    \gamma_{\bar{\hat{m}}\bar{\hat{n}}\bar{\hat{p}}}\eta_+=e^{-\mc{A}+i\bar{\mc{B}}}\bar{\Omega}_{\bar{\hat{m}}\bar{\hat{n}}\bar{\hat{p}}}\eta_{-}\,.
\end{equation}
Finally, from the fact that $\eta_-$, $\gamma_{\bar{\hat{m}}}\eta_-$, $\eta_+$, $\gamma_{\hat{m}}\eta_+$ are linear independent, we derive the two BPS conditions
\begin{equation}
    \d_{[\hat{\mu}}X^{\hat{m}} \d_{\hat{\nu}]}X^{\bar{\hat{n}}}J_{\hat{m}\bar{\hat{n}}}=0\,,
\end{equation}
\begin{equation}
    \d_{\hat{\mu}}X^{\hat{m}}\d_{\hat{\nu}}X^{\hat{n}}\d_{\hat{\rho}}X^{\hat{p}}\Omega_{\hat{m}\hat{n}\hat{p}}=-i e^{-2i\theta}e^{\mc{A}-i\mc{B}}\sqrt{-h}\epsilon_{\hat{\mu}\hat{\nu}\hat{\rho}}\,.
\end{equation}
Multiplying both sides with 
\begin{equation}
    \int_{\mc{C}}\dd^3 x\epsilon^{\hat{\mu}\hat{\nu}\hat{\rho}}\,,
\end{equation}
we find the relation with the holomorphic volume $\mc{V}(\mc{C})$ of the 3-cycle $\mc{C}$:
\begin{equation}
    \mc{V}(\mc{C})=\int_{\mc{C}}\dd^3 x\sqrt{-h}=i e^{2i\theta}e^{i\mc{B}}e^{-\mc{A}}\int_{\mc{C}}\Omega\,.
\end{equation}
By choosing a suitable phase factor, we obtain the final result
\begin{equation}
    \mc{V}(\mc{C})=e^{-\mc{A}}\left|\int_{\mc{C}}\Omega\right|=\sqrt{\frac{8\mc{V}_0}{\mc{V}_{\Omega}}}\left|\int_{\mc{C}}\Omega\right|\,.
\end{equation}

\section{Conifold singularity on mirror Calabi-Yau threefolds}
\label{sec:conifold-sing}

In this section, we explain the details of the conifold singularity on Calabi-Yau threefolds with large $h^{1,1}$ and small $h^{2,1}$.

\subsection{Mirror quintic with $(h^{1,1},h^{2,1})=(101,1)$}

We start with an example of the quintic threefold $X$ with $(h^{1,1}(X),h^{2,1}(X))=(1,101)$, in the special form:
\be
z_1^5+z_2^5+z_3^5+z_4^5+z_5^5-5\psi z_1 z_2 z_3 z_4 z_5=0\,,
\ee
where $[z_1:z_2:z_3:z_4:z_5]$ are projective coordinates of $\mb{P}^4$. Note that the equation is invariant under the $G=\mb{Z}_5^3$ quotient action:
\be
[z_1:z_2:z_3:z_4:z_5]\sim [z_1:\omega^{\lambda_2}z_2:\omega^{\lambda_3}z_3:\omega^{\lambda_4}z_4:\omega^{\lambda_5}z_5]\,,\quad \left(\lambda_2+\lambda_3+\lambda_4+\lambda_5\equiv 0\ (\mathrm{mod}\ 5)\right)\,.
\ee
In this case we define $\omega=e^{\frac{2\pi i}{5}}$.

Consider the quotient space $X/G$, which has quotient singularities at the fixed loci under the $G$ action. For generic values of the parameter $\psi\neq \omega^k$ $(k\in\mb{Z})$, after a crepant resolution of $X/G$, we obtain the smooth mirror Calabi-Yau threefold $\widetilde{X/G}\equiv \mc{X}_6$ with $(h^{1,1},h^{2,1})=(101,1)$~\cite{Candelas:1990rm}.

Now if $\psi$ is a 5-th root of unity, there would be 125 conifold singularities on $X$. Without loss of generality, we assign $\psi=1$, and working in the patch of $z_1=1$ in the later discussions. The 125 conifold singularities on $X$ locate at
\be
\label{quintic-conifolds}
(z_2,z_3,z_4,z_5)=(\omega^{\lambda_2},\omega^{\lambda_3},\omega^{\lambda_4},\omega^{\lambda_5})\,,\quad (\lambda_2+\lambda_3+\lambda_4+\lambda_5\equiv 0\ (\mathrm{mod}\ 5))\,.
\ee
For example, to see the conifold singularity equation around $(z_2,z_3,z_4,z_5)=(1,1,1,1)$, we define $y_i=z_i-1$ $(i=2,3,4,5)$ and expand $X$ around $y_i=0$:
\be
10(y_2^2+y_3^2+y_4^2+y_5^2)-5(y_2 y_3+y_2 y_4+y_2 y_5+y_3 y_4+y_3 y_5+y_4 y_5)+\mc{O}(y^3)=0\,.
\ee
Now after taking the quotient $X/G$, one finds that the 125 singular points in (\ref{quintic-conifolds}) are exactly identified to one singular point $p:(z_2,z_3,z_4,z_5)=(1,1,1,1)$ on $X/G$. Moreover, as the point $p$ is away from any fixed locus of the $G$ action, we see that $p$ on $X/G$ is still a conifold singularity. 

In this case, one can first resolve the quotient singularities on $X/G$ in the same way as the case with a generic parameter $\psi\neq \omega^k$ $(k\in\mb{Z})$, to get a space $\widetilde{X/G}$ with the conifold singularity at $p$. Then one can perform a small resolution on the conifold singularity, to get the smooth $\widetilde{\mc{X}}_6$. As we have already fixed the single complex parameter $\psi=1$, $\widetilde{\mc{X}}_6$ is now rigid with no complex structure moduli, and we have $h^{2,1}(\tilde{\mc{X}}_6)=0$. Note that in Section 6 of \cite{Candelas:1990rm} it was commented that such $\tilde{\mc{X}}_6$ is not a K\"{a}hler manifold any more, due to that there is only $n_{\mc{C}}=1$ 3-cycle shrinking to zero volume at the conifold point, and there is no further linear relation, $n_L=0$. This can be verified from the computation of the prepotential near the conifold point, see e.g. \cite{Blumenhagen:2016bfp} and (3.9) of \cite{Alvarez-Garcia:2020pxd}.

\subsection{CY3 with $(h^{1,1},h^{2,1})=(99,3)$}
\label{sec:99-3}

Now we review the case of the main example in this paper, which is the CY3 $\mc{X}_6$ with $(h^{1,1}(\mc{X}_6),h^{2,1}(\mc{X}_6))=(99,3)$, following~\cite{Demirtas:2020ffz}.

We start from $X$ with $(h^{1,1}(X),h^{2,1}(X))=(3,99)$ described by the equation
\be
\label{Px-99}
P(x)= \prod_{i=1}^7 z_i-\psi_1 z_1^6-z_2^2-z_4^6 z_6^{6} z_7^6- z_3^6 z_4^{12} z_7^6-z_5^3 z_6^6 z_7^3-\psi_6 z_3^6 z_5^6-\psi_7 z_3^6 z_4^6 z_5^3 z_7^3\,.
\ee
$z_i$ are projective coordinates of a 4d favorable reflexive polytope with rays $v_i$ in the $i$-th column of
\be
\bp -1 & 1 & -1 & -1 & -1 & -1 & -1\\
3 & -1 & 0 & 0 & 0 & 0 & 0\\
-2 & 0 & 0 & 0 & 1 & 2 & 1\\
-1 & 0 & 1 & 0 & 1 & 0 & 0
\ep\,.
\ee
After an FRST of the 4d polytope, each column of the above matrix corresponds to a compact divisor $D_i:z_i=0$ of the smooth anticanonical hypersurface $X$. We choose a basis for the Picard group of $X$:
\be
H^1=D_6\ ,\ H^2=D_1\ ,\ H^3=D_1+D_3\,,
\ee
such that the non-vanishing triple intersection numbers among $H^i$ are
\be
\mc{K}_{333}=\mc{K}_{233}=\mc{K}_{223}=4\ ,\ \mc{K}_{133}=\mc{K}_{123}=\mc{K}_{122}=2\ ,\ \mc{K}_{222}=3\,.
\ee
The equation (\ref{Px-99}) is invariant under the $G=\mb{Z}_6\times\mb{Z}_6$ action:
\be
\ba
&(z_1,z_2,z_3,z_4,z_5,z_6,z_7)\sim (z_1,z_2\lambda_1^3,z_3,z_4,z_5\lambda_1,z_6\lambda_1,z_7\lambda_1)\cr
&(z_1,z_2,z_3,z_4,z_5,z_6,z_7)\sim (z_1,z_2,z_3,z_4,z_5\lambda_2^{-1},z_6,z_7\lambda_2)\,,
\ea
\ee
where $\lambda_1$ and $\lambda_2$ are 6th roots of unity. 

For generic values of the complex parameters $\psi_1$, $\psi_6$ and $\psi_7$, we can take the $G$ quotient action and define $X/G$. After the crepant resolution of quotient singularities on $X/G$, we obtain a smooth CY3 $\widetilde{X/G}\equiv \mc{X}_6$ that is the mirror of $X$, with $(h^{1,1}(\widetilde{X/G}),h^{2,1}(\widetilde{X/G}))=(99,3)$.

To see the conifold singularity, we would work in the patch of $z_3=z_5=z_7=1$. The solution to the equations
\be
P=\frac{\ptl P}{\ptl z_1}=\frac{\ptl P}{\ptl z_2}=\frac{\ptl P}{\ptl z_4}=\frac{\ptl P}{\ptl z_6}=0\,
\ee
happens at the special codimension-one subspace of the complex structure moduli space
\be
\label{psi7-conicond}
\psi_7=1+\psi_6\,,
\ee
and the 36 points
\be
z_1=z_2=0\ ,\ z_4^6=-1\ ,\ z_6^6=1-\psi_6\,.
\ee

We then convert $\psi_i$ to gauge invariant coordinates
\be
\tilde{\psi}_1=\frac{\psi_6}{\psi_7^2}\ ,\ \tilde{\psi}_2=\psi_1\ ,\ \tilde{\psi}_3=\psi_7\,,
\ee
and the LCS coordinates mirror-dual to curve volumes
\be
Z^a=\frac{\ln(\tilde{\psi}_a)}{2\pi i}+\sum_{\vec{n}\in\mathbb{N}_0^3}\alpha_{\vec{n}}^a\prod_{i=1}^3\tilde{\psi}_b^{n_b}\,.
\ee
Such LCS coordinates $(Z^1,Z^2,Z^3)$ should be identified with $(U^1,U^2,Z)$ in the main text. With such coordinates, the conifold locus is at $Z^3\equiv Z=0$, as expected.


Finally, after the quotient action $G$, the orientifold action quotient singularities, the 36 conifold points should be identified as a single conifold~\cite{Demirtas:2020ffz}.

After resolving all the quotient singularities and the conifold singularity at $p$, we arrive at the smooth Calabi-Yau threefold $\tilde{\mc{X}}_6$. Such a CY3 only has two complex structure moduli, due to the constraint $(\ref{psi7-conicond})$ to achieve the conifold singularity. In this case, as computed from the GV invariants~\cite{Demirtas:2020ffz}, there are $n_{\mc{C}}=2$ vanishing 3-cycles at the conifold point and $n_L=1$ relation between them. In this case as there is an even number of vanishing cycles, the new Calabi-Yau threefold after the conifold transition is still K\"{a}hler, according to \cite{friedman1986simultaneous} and Theorem 2.6 of \cite{Friedman:2024zid}.

\bibliographystyle{JHEP}
\bibliography{biblio.bib}

\end{document}